\documentclass[smallextended]{svjour3}       % onecolumn (second format)

\usepackage{graphicx}
\usepackage{amsmath,amssymb,amsfonts}
\usepackage{algorithmic}
\usepackage{graphicx}
\usepackage{textcomp}
\usepackage{xcolor}
\usepackage{soul}
\usepackage{booktabs}
\usepackage[]{mdframed}
\usepackage{makecell}
\usepackage{hyperref}
\usepackage{multirow}
\usepackage{tcolorbox}
\usepackage{listings}
\usepackage{color}
\usepackage{algorithm}
\usepackage{multicol}
\usepackage{enumitem}
\usepackage{array}

\usepackage[authoryear]{natbib}
\definecolor{darkgreen}{rgb}{0,0.5,0}
\renewcommand{\thetable}{\arabic{table}}
\begin{document}

\title{Detection, Classification and Prevalence of Self-Admitted Aging Debt
}

\author{Murali Sridharan \and Mika Mäntylä \and Leevi Rantala}


\institute{Murali Sridharan \at
              University of Oulu \\
              \email{murali.sridharan@oulu.fi}           %  \\
           \and
           Mika Mäntylä \at
             University of Helsinki \\
              \email{mika.mantyla@helsinki.fi}
           \and
           Leevi Rantala \at
             University of Oulu \\
              \email{leevi.rantala@oulu.fi}              
}

\date{Received: date / Accepted: date}

\maketitle

\begin{abstract}

\textbf{Context:} Previous research on software aging is limited with focus on dynamic runtime indicators like memory and performance, often neglecting evolutionary indicators like source code comments and narrowly examining legacy issues within the TD context. %Previous research on software aging primarily focused on dynamic runtime indicators like memory and performance metrics, often neglecting static indicators like source code comments. Additionally, these studies typically viewed legacy issues as part of technical debt, which may restrict the broader understanding of software aging.

%This study explores software aging from the aging decay dimension, focusing on how software deteriorates over time. 
\textbf{Objective:} We introduce the concept of Aging Debt (AD), representing the increased maintenance efforts and costs needed to keep software updated. We study AD through Self-Admitted Aging Debt (SAAD) observed in source code comments left by software developers.

\textbf{Method:} We employ a mixed-methods approach, combining qualitative and quantitative analyses to detect and measure AD in software. This includes framing SAAD patterns from the source code comments after analysing the source code context, then utilizing the SAAD patterns to detect SAAD comments. In the process, we develop a taxonomy for SAAD that reflects the temporal aging of software and its associated debt. Then we utilize the taxonomy to quantify the different types of AD prevalent in OSS repositories. %iterative feature bootstrapping and pattern recognition methods to detect SAAD and we develop a taxonomy for SAAD that  reflects the temporal aging of software. %We employ a mixed-methods strategy that combines qualitative and quantitative methodologies to analyze AD. This study begins with qualitative analysis to detect and classify features of Aging Debt from source code comments, followed by quantitative analysis to assess their prevalence in software repositories. Iterative feature bootstrapping and simple pattern recognition methods are utilized to detect AD in source code comments. We, then establish a taxonomy for categorizing software aging and AD.

\textbf{Results:} Our proposed taxonomy categorizes temporal software aging into Active and Dormant types. Our extensive analysis of over 9,000+ Open Source Software (OSS) repositories reveals that more than 21\% repositories exhibit signs of SAAD as observed from our gold standard SAAD dataset. Notably, Dormant AD emerges as the predominant category, highlighting a critical but often overlooked aspect of software maintenance. %Analysis of 9000+  repositories reveals that over 19\% exhibit SAAD, with Dormant AD being the most common. 

\textbf{Conclusion} As software volume grows annually, so do evolutionary aging and maintenance challenges; our proposed taxonomy can aid researchers in detailed software aging studies and help practitioners develop improved and proactive maintenance strategies. %We identified 78 features indicative of AD and established a taxonomy categorizing it into Active and Dormant types. Analysis of 9,000+ Open Source Software (OSS) Java repositories in the PENTACET corpus reveals that over 19\% exhibit Aging Debt (AD), with Dormant AD being the most common. As software volume grows annually, so do aging and maintenance challenges; our taxonomy aids researchers in detailed software aging studies and helps practitioners develop improved maintenance strategies. %As the amount of software grows each year, so do the challenges related to software aging and maintenance. Our taxonomy for software aging helps researchers to investigate software aging in greater detail. It can also assist practitioners in designing enhanced maintenance strategies.
%Conclusions

%\subclass{MSC code1 \and MSC code2 \and more}
\end{abstract}

\section{Introduction}
\label{intro}
%Software aging is a serious software maintainability and development problem. 

\begin{center}
\emph{
``The last woe, and sometimes the last straw, is that the product over which one has labored so long appears to be obsolete upon (or before) completion.''  \citep{brooksjr1975}}
\end{center}
Software systems inevitably age and result in a host of maintenance-related challenges as they evolve. This aging is not just the result of software becoming outdated over time, but is intricately linked with the technological landscape and user needs \citep{parnas1994software}. For example, the shift from desktop-based applications to web and digital interfaces shows how technological advancements can make optimally functional systems at a certain point in time face compatibility issues and, at times, even obsolescence. An illustrative example is that of a UK nuclear submarine operating on Windows 98\footnote{\url{https://cnduk.org/trident-windows-xp-need-know/}} which highlights how technological advancements can render systems obsolete, although they remain in critical operations. Moreover, software aging is a crucial indicator of security vulnerabilities, as outdated components often carry unpatched vulnerabilities \citep{wang2019detecting}. The Knight Capital incident, where outdated software code caused a financial loss of 460 million US dollars in less than an hour\footnote{\url{https://www.bugsnag.com/blog/bug-day-460m-loss/}}, emphasizes the importance of proactive monitoring and the need for addressing software aging issues during software evolution. Therefore, understanding the multifaceted nature of software aging is crucial for addressing the maintenance-related challenges associated with them.

%Software aging is influenced by both intrinsic and extrinsic factors. Intrinsic factors include rapid aging from software faults \citep{huang1995software, grottke2022aging} and slower aging due to changes in regulations, technology, and maintenance \citep{parnas1994software, grottke2022aging}. Extrinsic factors, especially environmental shifts, can also hasten software aging \citep{wu2015software, costa2023software}. 

%The multifaceted nature of software aging that affects the performance and security of software systems can be assessed through both direct indicators, such as system-wide metrics, also known as aging indicators \citep{grottke2008fundamentals}, and indirect indicators, like dependability attributes such as availability and performability, that serve as proxies to gauge and measure the effects of aging \citep{pietrantuono2020survey}. Intrinsic factors influencing software aging are categorized into fast aging \citep{grottke2022aging}, which encompasses software faults \citep{huang1995software}, and slow aging, which encompasses change-induced software entropy \citep{parnas1994software} due to regulatory or policy changes, technology transitions, and maintenance activities \citep{grottke2022aging}. Extrinsic factors, especially environmental shifts, can also hasten software aging \citep{wu2015software, costa2023software}. 

%Rapidly evolving software systems can lead to deterioration where once optimal code becomes outdated. This degradation manifests in various forms: 
Source code can age in many ways, including code becoming incompatible, stale, deprecated, or obsolete. Incompatible code fails to function due to updated user requirements, libraries, or system environments, leading to discrepancies in expected functionalities. Stale code is functional but is no longer updated, whereas deprecated code is temporarily maintained before shifting to newer alternatives. In addition, the technological advancements and the changing requirements often render functioning code obsolete. Each type of these source code aging poses unique challenges from a maintenance perspective. In some cases, explicit workarounds are required to keep legacy code compatible with new features. We refer to these aging-related degradation and the ongoing efforts to keep the code functional and current as AD.   
%AD represents the cumulative challenges of maintaining software as it continuously evolves, encapsulating the effort needed to adapt and incorporate ongoing changes and technological advancements.

%Software systems evolve rapidly, causing previously optimal code to become outdated. Such deterioration of source code over time manifests in various forms, such as incompatible code that fail to work because of outdated or evolved user requirements, libraries, or system environments, leading to discrepancies in expected interactions or functionalities; stale code, which remains functional but has not been updated recently; deprecated code, which is functional but officially discouraged in favor of more efficient alternatives; and obsolete code, rendered irrelevant by changes in requirements or technological advancements. In some cases, explicit workarounds may be needed in the source code to make legacy code compatible with new features. We propose to observe this temporal degradation of the software and the consequent fixes in the source code that address software aging, referring to it as \textbf{Aging Debt (AD)}. It refers to the cumulative challenges associated with maintaining software as it evolves and ages over time. It encompasses the effort required to sustain the software as it undergoes continuous changes and incorporates technical advancements.

In this study, we explore  source code comments in the software to understand how software ages from the perspective of developers. By analyzing these comments along with the code context, we aim to uncover details about parts of the code that are deteriorating, components that have become outdated, and how software evolves and requires updates. Contextually analyzed comments that explicitly or implicitly express concerns or fixes related to the software's aging are classified as SAAD comments. Subsequently, we conduct an in-depth analysis of the features (i.e., SAAD patterns) to develop a comprehensive taxonomy. Then, we utilize the taxonomy to assess the prevalence of the different types of AD in OSS repositories.
%In this study, we delve into Aging Debt (AD), %specifically from the source code. This exploration is conducted 
%by analyzing source code comments containing annotations left by developers that provide perspectives on the software's aging process. 
%Comments explicitly or implicitly indicating concerns/fixes related to the software's aging are classified as \textbf{Self-Admitted Aging Debt} after analysing the code context. %To the best of our knowledge, this is the first work that systematically studies software aging through the source code comments left by developers. 
%SAAD comments offer an insight into the developers' recognition of aging factors impacting software components, offering insights into the maintenance challenges posed by AD. Subsequently, we conduct an in-depth analysis of the features (i.e., SAAD patterns) to develop a comprehensive taxonomy on them. Utilizing this taxonomy, we then assess the occurrence of AD within various Open Source Software (OSS) repositories, aiming to provide an overview of SAAD's presence. % and impact. %ultimately providing a holistic understanding of the Self-Admitted Aging Debt (SAAD).

Examples of SAAD include:
%\begin{quote} 
\begin{lstlisting}
// Not used;

// deprecated in MySQL 5.7.11 and MySQL 8.0.0.

// this macro is not needed in ES6

// These are outdated but we'll probably keep them 
// forever anyway for backwards compatibility.
\end{lstlisting}
%\end{quote}

%/**
%	 * This class is only used statically so a public 
%  * constructor is not needed.
%	 */

To the best of our knowledge, this is the first effort to explore and detect software aging through source code comments and to delineate the concept of SAAD. This work offers a systematic approach to identify aging-related debt through source code comments and enables software teams to explicitly track and focus AD instances. With such explicit focus, AD instances can receive their due priority from a maintenance perspective, which are often overlooked or implicitly treated as TD.

\begin{tcolorbox}
\textbf{Contributions of this paper:}
%\begin{itemize}
%\item 

%\textbf{1.} We introduce Aging Debt as a metaphorical representation of pending maintenance effort due to temporal aging.
%\item 

\textbf{1.} We propose Self-Admitted Aging Debt (SAAD) as Aging debt (AD) appearing in source code comments, and we empirically assess the novel approach of detecting software aging through source code comments.
%\item 

\textbf{2.} We develop a taxonomy for SAAD encompassing various aspects of software aging.

\textbf{3.} We share the manually verified gold standard dataset comprising 2,562 SAAD comments along with their code context. Additionally, we share the manually reviewed 145 software aging-related features and 399 SAAD patterns used in our study to support further research and ensure the replicability of our findings.
%develop a taxonomy for Self-Admitted Aging Debt encompassing various aspects of software aging from a software maintenance perspective.

%\textbf{3.} We share the dedicated dataset of Self-Admitted Aging Debt afflicted source code comments along  with manually reviewed SAAD features \footnote{10.5281/zenodo.11183044}, to support further research and ensure the replicability of our findings.
%\end{itemize}
%\end{mdframed}
\end{tcolorbox}

%SAAD highlights code fragments that decay or misalign with current technologies and methodologies, necessitating modifications. 
%Our research focuses on exploring software aging manifestations within source code comments. We gather textual features related to Aging Debt (AD) indicators. 
 %It empowers practitioners with actionable insights, facilitating more informed decision-making in the face of software aging.

In the forthcoming sections, we present our methodological framework. Section \ref{foundational_context} describes in detail the concepts attributing to AD and elucidates its relation to TD. The design of our study,
%mixed-methods research methodology, tailored for our study's intricacy, 
is described in Section \ref{rd}. Our data sources are summarized in Section \ref{data_section}. We explore source code comments for SAAD identification in Section \ref{aad_exploration}, categorize detected SAAD in Section \ref{aging_taxonomy}, and assess its prevalence in Section \ref{aging_prevalence}.  Past research on software aging is covered in Section \ref{rw}, followed by validity considerations in Section \ref{validity} and conclusions in Section \ref{conclusion}.

\section{Theoretical Background}
\label{foundational_context}
In this section, we introduce terms and concepts that are related to and partially overlap with our focus on SAAD. This discussion demonstrates how our work aligns with and builds upon existing research.

\subsection{Two Views of Software Aging}
\label{sec:two-views}
On a high level, software can be observed from two alternative perspectives which are named depending on the origin, as outlined in Table \ref{tab:sw_views}. In software architecture research, they are called static and dynamic views \citep{riva2002combining}. Levels closer to implementation refer to these views as compile-time and runtime \citep{baralis1998compile}. This distinction is also available in software quality terminology \citep{159342}, where ``bugs'' in code are called ``faults'', while ``bugs'' that occur during runtime are called ``failures''.

%Dependeability, Reliability, Dynamic, Runtime, Aging
%Evolutionary, Maintenance, Static, Compile time  

When it comes to software aging, we propose that the first view connects software aging to quality attributes of software maintainability and evolvability, while the second view connects software aging to reliability and dependability. Past work has extensively studied software aging in the field of software reliability \citep{garg1998methodology, castelli2001proactive, cassidy2002advanced, huo2018using,liu2021cloud,andrade2021comparative, grottke2008fundamentals}. 
In software reliability research, software aging refers to the phenomenon that, ``\textit{as the runtime period of the system or process increases, its failure rate also increases}'' \citep{grottke2008fundamentals}. This type of aging has been studied with methods such as investigating memory leaks, CPU spikes, and system workload. Some works have adapted the defect prediction approach to software runtime aging and used code metrics to predict software failures \citep{cotroneo2013predicting, liang2021within}. Dynamic aging indicators, as seen in Figure \ref{aging_position}, fluctuate during system operation due to user interaction, system load, and environmental conditions.

Finally, a recent study \citep{grottke2022aging} categorized software aging into slow and fast, reflecting the rate at which the aging process progresses. Here, slow aging reflects aging that takes years to accumulate, i.e., the one occurring during software evolution, while fast aging refers to runtime aging.

Throughout the rest of the paper, View 1 shall be referred to as Evolutionary Aging, while View 2 shall be referred to as Runtime Aging. These two do not originate from the same rows as observed in Table \ref{tab:sw_views}, but we have reasons for these choices. Evolutionary aging is chosen because it connects this view to the decades of research done in the area of software evolution and maintenance. A logical pair for evolutionary aging would be reliability aging; however, Evolutionary Aging can also cause reliability issues, for example, using an outdated library. Therefore, view 2 shall be called Runtime Aging, which clearly distinguishes it from Evolutionary Aging.

In this paper, we study evolutionary aging with an explicit focus on source code comments to detect software aging. Evolutionary aging also includes version histories and lagging dependencies (known as Technical Lag). Evolutionary indicators, as seen in Figure \ref{aging_position}, remain constant unless deliberately altered and are typically associated with the source code or development artifacts.

\begin{table}
    \centering
    \caption{Two views of software from different origins}
    \begin{tabular}{ccc}
    \toprule 
         Origin & View 1 - Evolutionary Aging & View 2 - Runtime Aging \\
         \midrule
         Architecture&  Static& Dynamic\\
         Code&  Compile-time& Runtime\\
         Bug&  Fault& Failure\\
         Quality&  Maintainability, Evolvability& Reliability, Dependability\\
 Speed& Slow&Fast\\
 \bottomrule
 & &\\
    \end{tabular}
    \label{tab:sw_views}
\end{table}

%In this paper we study XX aging that receive

%We propose a different perspective, distinguishing software aging based on the context of its occurrence: evolutionary (runtime) and static, which are especially relevant from a development standpoint. This distinction emphasizes where and how aging can be observed and managed, directly linking to development artifacts/practices. %argue that there are two views of software aging evolutionary (runtime) and static based on their context of occurrence which is particularly relevant from a development perspective emphasizing where and how aging can be observed and managed, linking directly to development practices. 

\subsection{Concepts related to Evolutionary Aging}

\begin{figure}[htbp]
  \centering
  \includegraphics[width=0.9\linewidth]{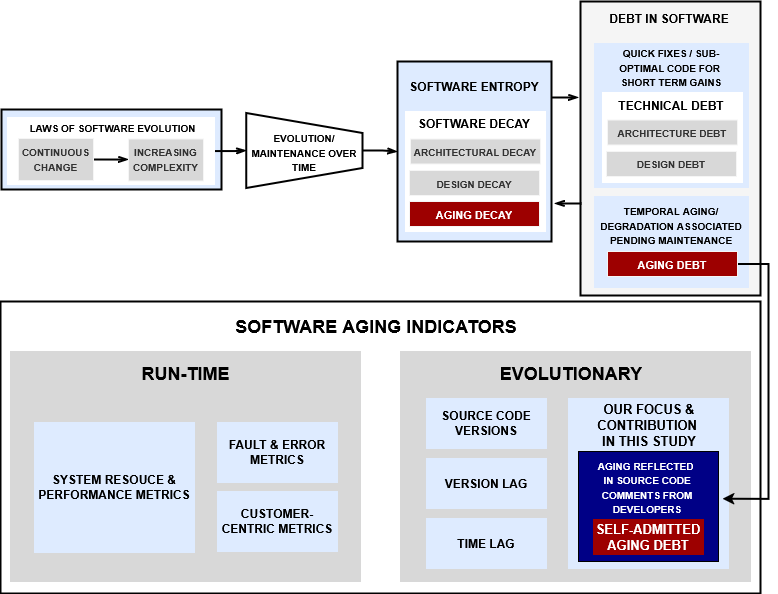}
  \caption{Software Aging Indicators and Factors}
  \label{aging_position}
\end{figure}

The concepts of Software Entropy, Software Decay (including Architectural Decay, Design Decay, and Aging Decay), TD (primarily comprising Architectural Debt, Design Debt, etc.,) and our proposed AD are intricately related. Figure \ref{aging_position} shows the intricate relationship between them and highlights the factors contributing to software aging, indicating our contribution in this work.

Software Entropy is the overarching concept encompassing the increase in complexity and disorder within a software system. High software entropy means that the system has become complex and less organized, which leads to software decay. Software Decay is the deterioration of the software's structure and quality over time. It includes Architectural Decay (decline in system architecture), Design Decay (erosion of design patterns), and Aging Decay (deteriorating software effectiveness and relevance over time). Each of these types of decay contributes to the overall increase in entropy within the software system. As the decay progresses, it can lead to increased entropy if not managed properly. TD and AD can contribute to both software entropy and software decay. As compromises are made through quick fixes and temporary solutions, they can increase the system's complexity (contributing to entropy) and lead to its deterioration (contributing to decay).

While substantial research has been dedicated to exploring Architectural Decay \citep{riaz2009architectural,hassaine2012advise,mo2013mapping,le2016relating,le2018empirical} and Design Decay \citep{izurieta2007software,izurieta2013multiple,dale2014impacts}, the concept of Aging Decay has received comparatively less attention. Our work focuses on Aging Decay and, by extension, AD. Aging Decay \citep{parnas1994software} centers on the temporal degradation of source code components. This degradation detrimentally affects the software's performance and adaptability to new requirements and technological standards. Concurrently, there is a hidden debt accrued with the compromises made in managing the temporal degradation and ensuing obsolescence of source code components, which we refer to as \textbf{`Aging Debt'}. AD represents the cumulative maintenance cost and the eventual need for system modernization, which left unattended, accrues over time as software components become progressively obsolete. It is characterized by the degradation of code quality and functionality resulting from failure to address the evolving technological landscape, changing user requirements, and lack of required feature updates. AD encapsulates the implicit costs and maintenance burden from the extended use of aging software components. %It encapsulates the implicit costs and technical challenges associated with the extended use of aging software components, ensuring that the system is updated and includes the effort required to clean or remove outdated code to ensure optimal performance.

The subsequent sections explain each of the contributing factor and the related concepts of AD.

\subsection{Laws of Software Evolution}
The first law of software evolution, known as Continuing Change, mandates that software must be continually adapted to maintain optimal performance; failing to adapt leads to degradation \citep{lehman1996laws}. Software, much like living organisms, evolves and expands, which necessitates ongoing modifications in response to environmental changes and external feedback. Neglecting this adaptation results in a gradual decline in user satisfaction over time that contributes to increased complexity. This need for continuous adaptation forms the basis for the second law of software evolution, Increasing Complexity \citep{lehman1996laws}. The complexity of software inherently increases as it evolves. 
If this growing complexity is not actively managed, the challenges of software maintenance escalates. With increasing complexity, the capacity of the software to grow diminishes due to the increased effort required for its management and maintenance. These factors highlights that as software evolves it invariably becomes more complex over time.

\subsection{Software Entropy}
Similar to the concept of entropy in Thermodynamics (disorder in physical systems) \citep{clausiusentropy} and Information Theory (uncertainty in communication messages) \citep{shannon1948mathematical, cropper1986rudolf}, software entropy \citep{berlinger1980information} measures the disorder and complexity in evolving software. It reflects the level of disorganization or randomness, which tends to increase over time, thereby increasing the overall complexity for maintenance-related activities. Such increasing complexity makes software difficult to understand or modify, limiting its maintainability and quality.

The idea of using entropy as a complexity measure in software was among the first adaptation from Information Theory \citep{berlinger1980information}. Later, several studies have employed entropy to quantify software complexity \citep{davis1988study, cook1990information, harrison1992entropy, cha1993complexity, kim1995complexity}. Entropy is also used to assess the impact of changes on software, considering systems as collections of various models (Requirements, Analysis, Design, Implementation) and measuring the ``Structural Information'' or the effort required to understand changes \citep{arnold1996software}. This concept is extended by a change entropy model that focuses on the unpredictability of changes made to software over time \citep{hanssen2009maintenance}. Change entropy is further extended to empirically assess the effects of refactoring, using the number of contributors, the involvement of source code components in design patterns, and commit notes on change entropy \citep{canfora2014changes}. Additionally, entropy is applied to software trace classification and bug prediction, further broadening its utility in understanding and managing software evolution \citep{miranskyy2012using, arora2019computation}.

Another study found that the more organized a software program is, the less entropy it entails, indicating less disorder \citep{mohanty1981entropy}. Further research found that higher entropy correlates with increased disorder within software, which led to the use of entropy as a metric for software maintenance \citep{chapin1989entropy}. Later, a Structure Model was developed to assess and maintain system quality, aiming to manage software's intrinsic entropy over time \citep{lanubile1992maintainability, lehman1989uncertainty}. In addition, entropy is applied to measure coupling and cohesion within modular systems \citep{allen1999measuring} and to evaluate the overall degradation of software \citep{bianchi2001evaluating}. Another study found that complex code increases software entropy \citep{hassan2003chaos}. Within Agile frameworks, the reduced maintainability is attributed to software entropy, suggesting a two-step approach to balance software entropy with agility \citep{hanssen2010software}. These studies position software entropy as a key indicator of system disorder.

%In this work, we view software entropy as a significant factor in contributing to and measuring software aging, building upon traditional perspectives.

\subsection{Software Decay}
Code decay refers to the deteriorated state observed in software beyond which further modifications necessitate a redesign \citep{karr1996empirical, porter1997fundamental}. Factors that cause code decay \citep{eick2001does} include inappropriate architecture, violations of original design principles, unclear requirements, time pressures that lead to shortcuts, inadequate programming tools, organizational issues (such as low morale or poor communication), programmer variability, and insufficient change management processes. Some of these factors, such as time pressures, unclear requirements, or inadequate programming tools may cause developers to introduce shortcuts that manifest as TD. Factors such as inappropriate architecture (i.e., architecture becoming outdated as the system evolves) and violations of original design principles (i.e., the initial design becoming outdated as new ones emerge) are potential indicators of temporal degradation in software. Although the introduction of quick and temporary solutions can also make the design and architecture outdated over time. Our focus and objective in this study is to utilize the explicitly available aging-related notes present in the source code comments to identify instances of temporal degradation of software resulting from software aging. To the best of our knowledge, this study is the first effort that studies software aging as observed in the source code comments.

%Factors that cause code decay \citep{eick2001does} include inappropriate architecture, violations of original design principles, unclear requirements, time pressures leading to shortcuts, inadequate programming tools, organizational issues like low morale or poor communication, programmer variability, and insufficient change management processes. Some factors of code decay such as time pressures, unclear requirements, or inadequate programming tools may cause the developers to introduce shortcuts manifesting as TD. While quick temporary fixes and patches introduced can also make the design and architecture becoming outdated over time. The instances without quick temporary hacks or patches are absolutely due to temporal degradation. The temporal degradation of software is often overlooked and to the best of our knowledge this work is the first attempt in studying software aging through source code comments. The element of Some factors like inappropriate architecture (architecture becoming outdated as system evolves over time), violations of original design principles (initial design becoming outdated as new ones emerge) are potential indicators of temporal degradation of software. The temporal aging and degradation of software is overlooked.  
%Code decay, a characteristic state of software, reflects the deterioration of code quality, maintainability, and adaptability. It is primarily 
 
\subsubsection{Architectural Decay}
Architectural decay reflects the architectural degradation of a software system over time. It involves the structure of the system, including components, their features, and interactions. Macro-architecture concerns the overall system layout, involving various architectural styles, while Micro-architecture focuses on the organisation of components and their interactions within specific design patterns \citep{garlan1993introduction}. Architectural decay occurs when the designed architecture deviates uncontrollably owing to the introduction of unplanned modules (usually from adding or modifying features or fixing defects) and their interactions \citep{lindvall2002avoiding}. Although architectural decay often requires insights from the original architects, early strategies to detect architectural decay involve reverse-engineering the actual architectural design from the source code to align it with original goals \citep{hochstein2003diagnosing, hochstein2005combating}.

\subsubsection{Design Decay}
Although related to architectural decay, design decay, specifically refers to the deterioration of design patterns over time. This relates to the compromise of structural integrity and intended functionality of design patterns \citep{izurieta2007software}. Grime \citep{izurieta2007software} refers to the accumulation of unnecessary or unrelated code within design pattern classes. It increases complexity making maintenance or enhancements challenging and contributing to the overall system decay.

\subsubsection{Aging Decay}
Aging decay refers to the gradual degradation of software due to the failure to update and adapt to new features/requirements, technologies, or environments \citep{parnas1994software}. This type of decay makes the software less relevant as it becomes outdated, incompatible with current standards, and inefficient to maintain or integrate with modern technologies \citep{bandi2013empirical}.

As newer technologies and programming paradigms emerge, they lead to the obsolescence of existing technologies, complicating maintenance. This is one of the primary factors of aging decay that affects the software's relevance and performance over time. It reduces the ability of the software to integrate with current systems and comply with current standards. Decreased compatibility poses challenges for older software to remain compatible, affecting its performance and functionality. The performance degradation resulting from evolving user expectations and advances in hardware capabilities further contributes to decay.

%The characterization of aging decay involves several key factors that impact the relevance and performance of a software system over time. Technological obsolescence is a primary factor that occurs as newer technologies and programming paradigms emerge. This obsolescence leads to the deprecation of older technologies and features, complicating maintenance and reducing the ability of the software to integrate with modern systems or comply with current technological standards. 

%Another significant issue is decreasing compatibility, which arises from continuous updates to operating systems, platforms, software and associated technology stacks. This evolution poses challenges for older software to remain compatible, adversely affecting its usability and functionality. Also, performance degradation plays crucial role. With evolving user expectations and hardware capabilities, adequately performing software may fall short of the expected standards for efficiency which further contributes to its decay. % As user expectations and hardware capabilities evolve, software that once performed adequately may no longer meet the expected standards for speed, responsiveness, or efficiency, further contributing to its perceived decay.

\subsection{Deprecation}
Software deprecation is defined as \textit{``the process of orderly migration away from and eventual removal of obsolete systems''} \citep{segoogle2020}. It indicates aged software that is either obsolete or no longer supported due to new alternatives. An experience report from Google highlights the need to move away from deprecated elements by highlighting the effects of using stable but soon-to-be deprecated technologies and adopting cutting-edge but incomplete ones \citep{avgeriou2016managing}.

Past research on deprecation has primarily focused on APIs \citep{robbes2012developers, zhou2016api, bonorden2022api}, build files \citep{morgenthaler2012searching}, and micro-service test bases \citep{sundelin2020hidden}, among others. The reasons for API feature deprecation in Java are critically examined using both manual and automated methods \citep{sawant2018features}. Another study investigates npm package release deprecation, highlighting their prevalence and impact on client packages \citep{cogo2021deprecation}. Tool-related efforts to mitigate the impacts of software deprecation include Diff-CatchUp \citep{xing2007api} and SemDiff \citep{dagenais2009semdiff}. Diff-CatchUp identifies API changes and suggests replacements for obsolete APIs. SemDiff recommends alternatives for deprecated or removed framework methods by analyzing the evolutionary adaptations of the framework. 

However, our study analyzes source code comments for aging-related information, offering a complementary view of software aging than deprecation alone. While existing works identify deprecation as code components that should be avoided or replaced, it overlooks the temporary maintenance effort associated with deprecated code and does not provide specific strategies for addressing the accumulated maintenance burden associated with aging code components.

%While past work on deprecation focused on APIs \citep{robbes2012developers, zhou2016api, bonorden2022api}, build files \citep{morgenthaler2012searching}, micro-services \citep{sundelin2020hidden}, etc., Focus of our study, source code comments containing aging information provide a broader picture of aging than deprecation. Moreover, while deprecation signals what should be avoided or replaced, it does not inherently provide solutions or strategies to address the broader accumulated debt or systemic issues of aging code components. %This situation calls for a broader spectrum of analysis to effectively address software aging-related issues, as acknowledged by software developers.  

\subsection{Technical Lag}
Technical lag \citep{gonzalez2017technical} refers to the gap between the versions of software packages used in a software system and the most recent versions available upstream (the official sources that distribute updates, including enhancements, fixes, and security patches). It measures the  difference between the deployed versions and the latest release from their sources. An empirical analysis of technical lag in npm package dependencies \citep{zerouali2018empirical} describes technical lag as time lag and version lag. Time lag measures the time lag between adopting a software version and the release of its newest version. Version lag denotes the difference between the deployed package version and the latest release. Their analysis shows that deployed package versions are often  behind the most recent release. Another study \citep{gonzalez2020characterizing} found that the delay in updating software from upstream to deployment depends significantly on the software development processes in upstream.

Technical lag provides a view of software aging by measuring package versioning. However, it cannot capture broader symptoms of software aging that are available in code comments, such as legacy compatibility issues or obsolete code. This highlights the need for a  comprehensive taxonomy of software aging.
%Technical lag \citep{gonzalez2017technical} refers to the gap between the versions of software packages used in a system and the most recent versions available upstream. %. 
%It indicates the discrepancy in software package versioning which emphasizes the difference between currently deployed versions and the latest available from their sources. An empirical analysis of technical lag in npm package dependencies \citep{zerouali2018empirical} describes technical lag as: time lag and version lag. Time lag measures the duration between adopting a software version and the release of its newest version. Version lag denotes the difference between the major, minor, or patch versions of the deployed software and the latest release. Their analysis found that deployed package versions are often lags behind the most current release. Another exploratory study \citep{gonzalez2020characterizing} suggests that the delay in updating software packages from upstream to deployment depends significantly on the software development processes in upstream.

%Technical lag provides a view on software aging by measuring the outdatedness of package versioning by examining versioning files. Thus it cannot capture broader symptoms of software aging that are available in code comments, such as obsolete code or legacy compatibility issues. This gap indicates a need for a more comprehensive taxonomy of software aging that includes, but is not limited to, technical lag and incorporates insights from source code comments left by software developers.

\subsection{Debt in Software}
In this subsection, we explore various types of debt in software, including TD, SATD, AD, and SAAD elaborating on the key differences between these concepts. 

\subsubsection{Technical Debt}
\label{sec:td_explanation}
TD, coined by \citep{cunningham1992wycash}, is a widely recognized metaphor in software maintenance. It refers to the costs and complexities added to software due to earlier expedient but sub-optimal development choices. This concept encapsulates various types of debt that arise when teams prioritize quick delivery over optimal code, leading to increased maintenance efforts in subsequent releases. 

Past research works identified different types of debt. Design debt \citep{guo2011portfolio,izurieta2013multiple} represents compromises of design principles. Architecture debt \citep{brown2010managing,kruchten2012technical} represents trade-off through architecture violations. Requirements debt \citep{kruchten2012technical} represents the gap between the actual requirements in the specification and the implementation owing to different constraints. Code debt \citep{bohnet2011monitoring} represents poor coding practices. Defect debt \citep{snipes2012defining} represents those known defects in the source code that are deferred due to constraints of time and resources. Test debt \citep{guo2011portfolio, codabux2013managing} represents sub-optimal testing practices. Documentation debt \citep{guo2011portfolio} represents inadequate documentation practices. Infrastructure debt \citep{seaman2013managing} represents inadequate or non-standard development tools, processes, and technology stack that impact development activities. People debt \citep{seaman2013managing} represents deficiencies and challenges within a software development team impacting productivity, efficiency, and team morale. Process Debt represents non-standard processes involved in software development activities. Build debt \citep{morgenthaler2012searching} represents inefficient build processes. Service Debt represents compromises made in software services (for example, web services, microservices, etc.,). Usability debt \citep{zazworka2014comparing} represents compromises made in the user interface or user experience (UI/UX) owing to time and resource constraints. Versioning debt \citep{greening2013release} represents shortfalls in the management and organization of source code versions (for example, unnecessary code forks, branches, etc.,).

\subsubsection{Self-Admitted Technical Debt (SATD)}
Self-Admitted Technical Debt (SATD) \citep{potdar2014exploratory} is coined in an exploratory study of source code comments in Java OSS repositories. It represents the intentional TD admitted by developers through source code comments. These comments are instances where developers explicitly acknowledge shortcomings in their code through comments. Examples of SATD: 
\begin{quote}
\begin{lstlisting}
// TODO: validate originateTime == requestTime.
/* FIXME: Handle other types */
\end{lstlisting}
\end{quote}
These comments act as markers or notes, indicating areas of the code that are known to require improvement, optimization, or refactoring, but for various reasons, such improvements have not been made at the time of development. One study identified and quantified five different types of SATD \citep{maldonado2015detecting}. Design debt: source code comments reflect design issues; Defect debt: source code comments reflect defects in the code; Documentation debt: source code comments that reflect poor documentation standards; Requirements debt: comments that reflect shortfalls/deviations from requirements/implementation; Test debt: comments that express inadequate tests. A recent study \citep{obrien202223} characterized 23 different types of machine learning related SATD in an empirical study including, model, datatype, configuration, dependency, interpretability, abstraction, etc.,

Despite extensive studies on various types of TD and SATD, research into the debt associated with aging decay remains limited. While some studies treat legacy as TD \citep{monaghan2020redefining}, the broader implications of aging-related decay remain largely unexplored.

\subsubsection{Aging Debt}
 %through states such as legacy, outdated, deprecated, stale, and obsolete. 
%These states reflect the temporal degradation of the software. % that begins when software is developed and deployed on a production server. %As software and the world around it evolves over time, this process advances, leading to the accumulation of refactorings at both granular and architectural levels. These refactorings are realized through various types of software maintenance, including corrective and adaptive maintenance. 
Software aging is a temporal process. During this aging process, what was once considered ideal source code may gradually transition from optimal to obsolete, reflecting the temporal degradation of software.

We define AD as the aggregate debt in terms of cost and effort associated with software upkeep throughout its temporal aging, ensuring its efficiency, and reliability. AD enables us to model the temporal aging of software comprehensively. By systematically identifying and categorizing temporal deterioration, AD addresses a critical risk that is often overlooked when narrowly focusing on resolving TD. 

Legacy code \citep{bennett1995legacy, ransom1998method} represents a crucial stage in software aging, highlighting the transition from its initial deployment state \citep{furnweger2016software}. TD is a prominent characteristic of legacy code, where code that is `good enough' is delivered to meet stringent deadlines, rather than striving for algorithmic perfection \citep{birchall2016re}. Other characteristics of legacy code include being untested and untestable, as well as inflexibility to accommodate new changes or features. This suggests that while legacy code may potentially contain TD, not all legacy code constitutes TD. Past research has established that although TD and legacy systems share some similarities, they are distinct concepts that require structured co-existence \citep{holvitie2016co}. Furthermore, a practitioner survey \citep{holvitie2018technical} highlights concerns about the risk of mislabeling legacy issues as TD, noting that many instances of TD originate from legacy systems. In addition, the study recommends that specific measures be taken to systematically transform complex legacy system issues into manageable TD instances, ensuring that both categories are properly identified and handled. On the other hand, the study by \citep{monaghan2020redefining} treats legacy issues as TD. This differing perspective between software legacy and TD is due to the lack of systematic analysis through the lens of software aging, i.e., aging software and its associated debt. 

A recent study \citep{monaghan2020redefining} captures practitioners' opinions on legacy software which includes: ``They tend to be older systems developed with older technology, typically older programming languages or even older hardware.'', ``The legacy system has been in place 10 or 15 years—it's at the end of its useful life now.'', ``I try not to use the terms `heritage' and `legacy' as they carry a connotation of old things that we shouldn’t have.''. While these statements exclusively highlight challenges associated with aging software, the researchers developed a new type of debt called ecosystem debt by redefining legacy from the perspective of TD. In contrast to their approach of examining legacy systems through the lens of TD and to overcome the perspective differences between legacy software and TD, our study systematically investigates software aging by analyzing source code comments left by developers. 

\subsubsection{Aging Debt versus Technical Debt}
\label{sec:AD_vs_TD}

TD \citep{cunningham1992wycash} often arises from deliberate choices developers make in response to environmental pressures such as deadlines, limited resources, or other urgent necessities. For example, developers might implement quick fixes before a product launch, neglecting comprehensive documentation, or ignoring coding standards to meet pressing deadlines. Additionally, TD can stem from a lack of skill or understanding in implementing certain parts of software efficiently.  

AD identifies debt associated with maintaining, refactoring, and evolving software as it ages over time. This includes compatibility with legacy code, managing obsolete or deprecated code, keeping up with updates or upgrades, and managing non-maintained or unused code that may require future cleanup or adaptation. 
  %Windows 1998, Windows 2003, Windows XP are examples of outdated software systems, primarily due to temporal degradation and evolution of technology. Some government agencies still use Windows XP \footnote{\url{https://www.techspot.com/news/100376-us-department-state-using-13-year-old-operating.html}}.  
%Aging Debt exclusively focuses on software aging related maintenance efforts, i.e., the due maintenance efforts required to keep the software optimal and current amidst evolving software requirements and business needs. It includes addressing legacy, stale or obsolete features, and functionalities, etc., in the source code.  
%While Aging Debt (AD) shares similarities with TD, it exclusively focuses on the maintenance required to keep software current, subtly accumulating over time, often not due to a conscious design decision but due to inevitable temporal degradation as technologies evolve and requirements change. For instance, a system built today with the latest stable software libraries does not have any AD, but it most likely has TD. However, over time, the system will become outdated, thereby incurring AD. Yet the TD will not change over time unless it is aging induced. 

\begin{figure}[htbp]
  \centering
  \includegraphics[width=0.6\linewidth]{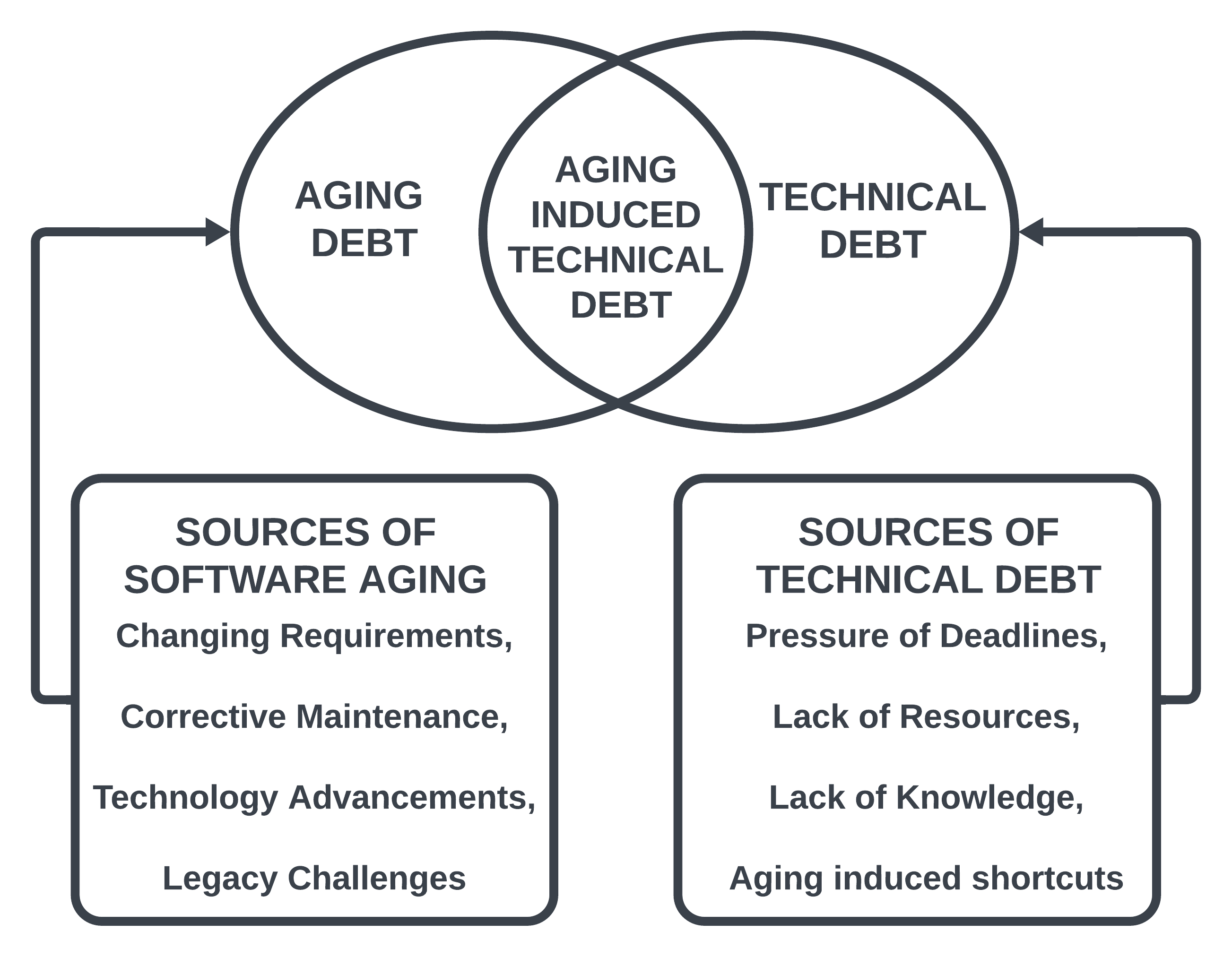}
  \caption{Aging Debt vs Technical Debt}
  \label{ad_td_relation}
\end{figure}

The relationship between AD and TD is illustrated in Figure \ref{ad_td_relation}. It shows how AD arises as software or its operating environment changes, failing to keep pace with evolving requirements or new technologies. This can lead to the persistence of legacy code that no longer meets current standards, the use of outdated software designs, or dependencies on obsolete libraries. 

Software evolution may lead to the accumulation of AD as older components become harder to maintain, often prompting developers to opt for quick fixes that result in TD, particularly under time or budget constraints \citep{sharma2018survey}. However, TD can also arise independently when quick fixes are implemented to meet imminent deadlines.  

The unique focus of AD on the evolutionary aging of source code provides a vital strategic dimension  beyond the traditional scope of TD. It offers a broad view of how software evolves over time, aligning with technological advancements, evolving dependencies, and changing requirements. %and shifts in the business context. Consequently, AD enables informed decisions that consider the broader implications of software evolution rather than merely addressing the immediate symptoms of poor design or rushed development typically reflected in TD.

The distinction extends to SAAD, which, like SATD \citep{potdar2014exploratory}, utilizes source code comments to identify debt. However, SAAD delves into the maintenance and evolutionary aspects of software aging, covering intentional and unintentional aspects. For example, code can become outdated, unused, or obsolete over time due to evolving technological landscapes or shifting user needs. While SAAD can arise without explicit intention or immediate project-specific needs, SATD results from intentional, short-term trade-offs during development and maintenance.

Consider the following code comment for example which reflects a workaround for the old code, demonstrating the intersection of AD and TD.
\begin{lstlisting} 
// Temporary workaround (for the old code).
\end{lstlisting}

The following example code comment exclusively reflects AD (pending maintenance effort due to temporal aging and subsequent degradation).
\begin{lstlisting} 
// Old code, needs rewrite
\end{lstlisting}

%Consider the following analogy that exemplifies the difference between AD and TD. Consider a car component installed quickly without proper functional and performance validations to meet a deadline, either knowingly or unknowingly, that it will incur future maintenance costs. This component represents TD because it involves a shortcut taken during rushed development, with deficiencies that are known and planned to be fixed later. Now, consider a similar component that, over time, has simply worn out or become less effective due to temporal aging, not because of a rushed or temporary fix but because it has reached the end of its useful life. Similarly, consider an enterprise application developed in Java 6 with little or no TD. As the ecosystem evolves, for example, with Java 8 and later introducing improved language features, developers will eventually need to refactor the application to take advantage of these enhancements. Such refactoring, driven solely by technological obsolescence (i.e., the temporal aging of software features and functionalities) rather than by any initial suboptimal design choices, exemplifies that AD can exist independently of TD. Since AD can exist even when TD is minimal because software inevitably ages and must be updated to remain efficient and compatible with evolving standards, 

We argue that by focusing explicitly on the software aging-related notes left in the comments by developers, we can isolate a subset of maintenance challenges that traditional TD overlooks and does not cover by definition (i.e., shipping expedient, suboptimal code to meet deadlines). %In principle, identifying software aging–related debt (i.e., AD) as TD may mislead practitioners by shifting focus from actual TD instances because TD typically arises from conscious decisions taken during development, whereas AD results from the inevitable temporal aging and deterioration of software systems.
In addition, software is a dynamic system that evolves over time \citep{lehman1996laws}. Since time is an independent factor in such a dynamically evolving system, temporal evolution related debt must be treated independently unlike other types of debt highlighted in Section \ref{sec:td_explanation}. Therefore, we posit that AD and TD are distinct and complementary views of debt in software, although they share some overlaps in symptoms and causes. Essentially, TD emphasizes trade-offs made for quick delivery, while AD emphasizes the temporal aging, degradation, and associated maintenance.

\section{Research Design}
\label{rd}
The goal of this research is to establish an initial understanding of AD as observed in source code comments, employing a pragmatic approach \citep{goldkuhl2012pragmatism}. Our pragmatic ontological and epistemological stances \citep{easterbrook2008selecting,maxwell2012qualitative,creswell2016qualitative} highlight dynamic and adaptable perceptions of reality, as well as the selection of methodologies based on practical utility. In our adoption of a pragmatic ontological position, we embrace a flexible understanding of AD, acknowledging its potential to evolve in response to practical implications and specific study contexts. On the epistemological front, our pragmatic stance ensures that we do not confine ourselves to predefined knowledge acquisition methods. Instead, we prioritize methodologies based on their tangible utility and efficacy in addressing our research objectives.

\begin{figure}[htbp] \centering \includegraphics[width=\linewidth]{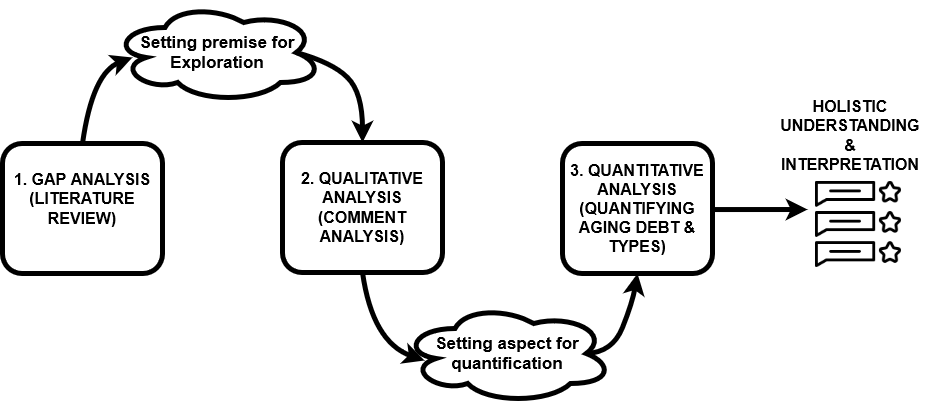} \caption{Research Design} \label{mm_design} \end{figure}

Adopting these pragmatic stances, our research employs a mixed-methods \citep{creswell1999mixed} approach characterized by a sequential exploratory design, combining qualitative and quantitative techniques to comprehensively understand AD within source code comments. Qualitative insights provide the detection and classification, while quantitative analysis measures prevalence, as shown in Figure~\ref{mm_design}.

Our approach recognizes that knowledge is actively constructed from the empirical evidence observed in source code comments, aiding in creating a practical taxonomy for managing AD facets. 
We address the following research questions:

\textbf{RQ1: How can SAAD be detected from source code comments using text features?}
We employ a systematic approach to identify relevant source code comments for AD detection. This process primarily involves Sense2Vec AI model for AD feature extrapolation. Then we analyse source code comments along with the source code context, employing the exploratory search framework. This framework begins with a very little information like seed features followed by an iterative querying (search the source code comments along with the code context) to identify code context evaluated SAAD patterns and followed by analysis. The analysis utilizes the code context evaluated SAAD patterns and perform pattern-based annotation of source code comments to detect AD. This annotation undergo multiple rounds of manual labeling of code comments with code context. In the process of manual labeling, we develop a fully manually verified gold standard SAAD comments dataset and code context evaluated SAAD pattern based machine annotated silver standard dataset for triangulation of the prevalence of AD for RQ 3. For a detailed explanation of these exploratory steps, please refer to Section~\ref{aad_exploration}. %Feature Bootstrapping method that refines our feature list and adds new AD features based on the source code comments data. These features are then used for automated annotation of source code comments for AD, which undergo multiple rounds of manual labeling to ensure quality. 

\textbf{RQ2: How can SAAD in source code comments be classified?}
Following the qualitative exploration of software aging detection in source code comments, the identified SAAD comments are qualitatively analyzed and grouped into categories based on the nature of AD, reflected in the source code comments. These categories are used to develop a taxonomy of AD categories iteratively. The developed taxonomy serves as a reference for the next quantitative phase. Section \ref{aging_taxonomy} lists the taxonomy creation steps in detail. 

\textbf{RQ3: What is the prevalence of SAAD in OSS repositories?}
In Section \ref{aging_prevalence}, we detail the steps for quantitatively evaluating AD prevalence in OSS repositories. This section marks a transition from the prior qualitative phase in which we explored and classified AD as part of RQ1 and RQ2. In this new quantitative phase, we assess the extent of AD and its various types in software repositories to determine the predominant form of AD in OSS repositories. The results from RQ1 (gold standard and silver standard SAAD dataset) and RQ2 (AD Taxonomy) provide the foundation for this quantification, focusing on identifying different AD types within source code comments. We utilize the silver standard dataset as a form of data triangulation \citep{denzin2012triangulation, flick2012can, da2013team} for evaluating RQ 3, in addition to the gold standard data, to assess if our findings are robust and generalizable beyond the gold standard dataset. 

For our study, we utilize the PENTACET \citep{sridharan2023pentacet} corpus, which comprises several million contextual natural language source code comments along with their code context mined from over 9,000 Java Open Source Software repositories. The authors of PENTACET corpus have employed relevant criteria to include most popular and active OSS repositories through git metadata such as star rating and frequency of commits. Overall, our research design aligns well with the principles of the sequential exploratory design, where the qualitative phase informs the development or refinement of instruments for the quantitative phase. Here, the initial qualitative exploration enables the development of AD taxonomy which later aids the quantification of different types of AD in OSS project repositories.

\section{Data}
\label{data_section}
In this study, we utilize the openly available PENTACET \citep{sridharan2023pentacet} corpus for exploring AD. Table \ref{tab:data_pentacet} contains the PENTACET data stats. The corpus has over 16 million natural language source code comments mined from over 9,000 OSS Java source code repositories. We chose the PENTACET corpus because the authors have captured the data recently, and it has several million contextual natural language (NL) source code comments with varying project size (Lines of Code), team size (based on the developers with commit access to the repository) along with several project and repository related attributes. The PENTACET authors have utilized \verb|SoCCMiner| \citep{sridharan2022soccminer} tool for mining the source code comments, and the mined data are clustered into four clusters, C1 through C4, based on the number of developers corresponding to the source code repository. 

%The authors of PENTACET corpus carefully included filtering criteria to include only active, popular OSS repositories. They excluded non-en (non-english), archived, tutorial-related repositories by checking for tutorial/assignment related keywords in the repository description. In addition, the PENTACET code comments, along with the code context, are retrieved from the latest and main branch of the repository, ensuring the data is a snapshot of the latest code when it was fetched.

\begin{table}[htbp]
  \begin{center}
  \caption{Data Stats}
  \label{tab:data_pentacet}
  \begin{tabular}{ccccc}
  
    \toprule
    \textbf{\thead {PENTACET \\ cluster }} & \textbf{\thead {\# of \\ Projects}} & \textbf{\thead{KLOC}}&\textbf{\thead{Team \\Size}} & \textbf{\thead{ \# of NL\\ comments}}\\
    \midrule
    1 &1,574 & 28,339 & 1 & 733,409\\
    2 & 2,202 & 48,355 & 2-3 & 1,766,513\\
    3 & 3,004 & 87,006 & 4-10 & 3,281,562\\
    4 & 2,314 & 289,783 & 11-30 &10,372,458\\
  \bottomrule
  \textbf{Total} & 9094 & 453,483 & & 16,153,942 \\
  \bottomrule
\end{tabular}
\end{center}
\end{table}

The PENTACET corpus \footnote{10.5281/zenodo.7757461} is available as a PostgreSQL \footnote{https://www.postgresql.org/} database. The authors of the PENTACET corpus included a host of filtering criteria to include only active and popular OSS repositories and excluded non-english, archived, tutorial-related repositories by checking for tutorial or university assignment-related keywords in the repository description. In addition, the code comments, along with the code context, are retrieved from the latest and main branch of the repository, which ensures the comments are from the latest versions when the corpus was created. We explore the AD phenomenon in the dataset with over 16 million natural language comments. The natural language (`NL') comments are identified using the \verb|NLON_STATUS| column from the \verb|COMMENT_ATTR| table of the PENTACET database.

%\section{Results \& Discussion}
\section{RQ1: SAAD Detection}
\label{aad_exploration}

\subsection{Motivation}

RQ1: How can SAAD be detected from source code comments using text features?

Our study emphasizes the utilization of source code comments as a mechanism for identifying SAAD. These readily available comments, created by developers for clarification or code documentation purposes, can act as pivotal indicators, highlighting sections of the code beginning to show aging signs. This enables the identification of specific code segments indicating software aging that may demand refactoring or even complete removal. By closely examining these developer comments using text patterns indicative of AD, we can effectively detect SAAD. % This examination reveals Aging Hotspots in the source code, which are vital for recognizing and addressing Aging Debt in software.  Here, we evaluate the following research question,

\subsection{Approach}
\label{sec:det_saad_approach}

Exploring a new phenomenon within such a large corpus of source code comments like PENTACET is both challenging and time-consuming. To manage this complexity, we utilize the Exploratory Search framework \citep{marchionini2006exploratory}, which recognizes that users often begin with a loosely defined information need (such as software aging-related text features) and refine their understanding through a combination of querying, browsing, and analysis. We adopt the same approach, first by identifying manually verified software aging-related features extrapolated from a set of seed features (i.e., core features). Then, querying and browsing form a cycle in which the PENTACET data corpus is queried using the software aging-related features, and the matched comments are further evaluated manually to extract SAAD patterns from those comments that reflect SAAD. The analysis stage involves utilizing the code context evaluated SAAD patterns for automated pattern-based annotation, and the matching comments are manually examined to determine whether a comment is truly an SAAD comment or merely an explanatory note. Figure \ref{exploring_aad} depicts the methodology adopted for exploring AD through source code comments within PENTACET \citep{sridharan2023pentacet} data corpus. The following steps explain the SAAD exploration approach in detail. Each step in Figure \ref{exploring_aad} is briefly explained as follows:

%\begin{figure*}
%  \centering
%  \includegraphics[width=0.9\linewidth]{exploring_ad.png}
%  \caption{Methodology: Exploring AD}
%  \label{exploring_aad}
%\end{figure*}

\begin{figure*}
  \centering
  \includegraphics[width=0.9\linewidth]{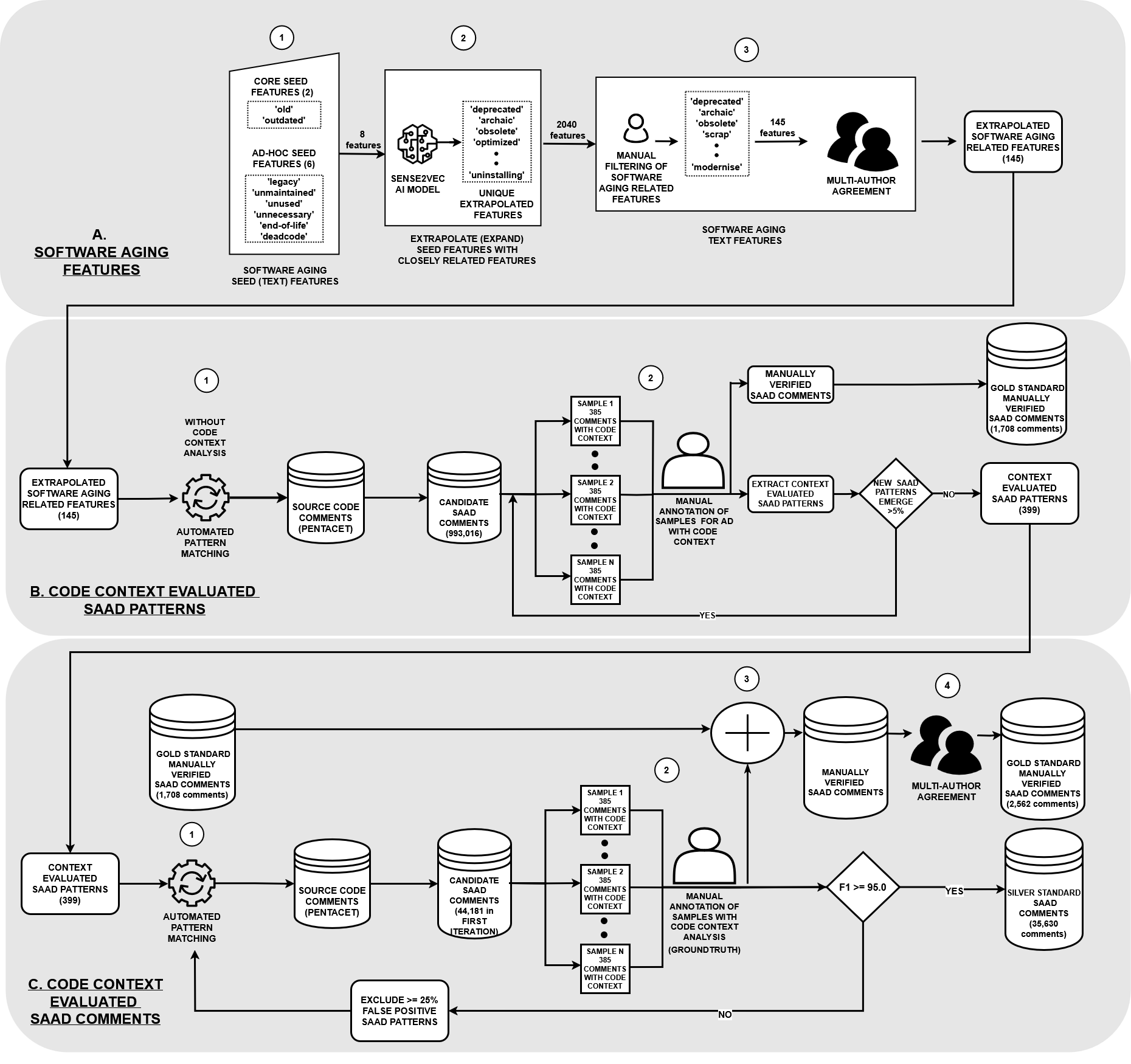}
  \caption{Methodology: Exploring SAAD}
  \label{exploring_aad}
\end{figure*}

%\begin{enumerate}[label=\Alph*.]
%\item 
\subsubsection{A. Software Aging Features} The focus of this step is to acquire an exhaustive set of manually verified software aging-related text features, see Figure \ref{exploring_aad} part A. The following steps explain in detail how the software aging-related text features are identified. It must be noted that these software aging features reflect software aging broadly without code context evaluation. We utilize this manually verified initial information (software aging features) later to extract SAAD patterns.
\begin{enumerate}[label=\arabic*.]
  \item \textbf{Seed Feature Identification:} Overall, we identify eight seed features (i.e., text keywords) related to software aging. These comprise two core features: `old' and `outdated' derived from the seminal paper on software aging \citep{parnas1994software}, and six ad-hoc features: `legacy', `unmaintained', `unused', `unnecessary', `end-of-life', and `deadcode'. 
  %based on the authors' experience during the development of the PENTACET dataset. 
  We arrived at the six software aging ad-hoc features that cover the theme of software aging, i.e., software becoming old, transitioning into `legacy', and being left `unmaintained'. This progression often results in the software becoming %outdated, 
  `unnecessary', or `unused', ultimately reaching its `end-of-life' and turning into `deadcode'.
  %We identify the core software aging features from the literature \citep{parnas1994software,matias2014systematic,araujo2011experimental,constantinides2009prolonging,bhasin2021software}. It provide our core text features, which consists of words ``old'', ``outdated'', ``version'', ``compatible'', ``legacy'', ``EOL''. 

  %TODO: Update figure 4 to include total 8 features 

  %TODO: update figure 5 to include only most accurate features

  %TODO: make 2.1 and 2.2 for 2 and 3

  %TODO: update 144 as 145 features

  %TODO: include an example of aging  feature and SAAD pattern

  %TODO: feature count in appendix to be corrected

  %TODO: make a table to differentiate feature from pattern

  %TODO: include the number of features in each step especially in software aging efature steps and SAAD pattern identification step.

      \item 
  
  \textbf{Feature Extrapolation:} After identifying the seed features, the next step is to detect software aging-related text features that are related to the seed features. To accomplish this, we use textual feature extrapolation, a process of generating additional related features based on the initial seed features. Figure \ref{sense2vec_feature_extrapolation} illustrates the theme of extrapolating features and identifying relevant software aging-related features. For this purpose, we leverage a well-established neural network model called Sense2Vec \citep{trask2015sense2vec}. 

  \begin{figure}[htbp]
  \centering
  \includegraphics[width=0.9\linewidth]{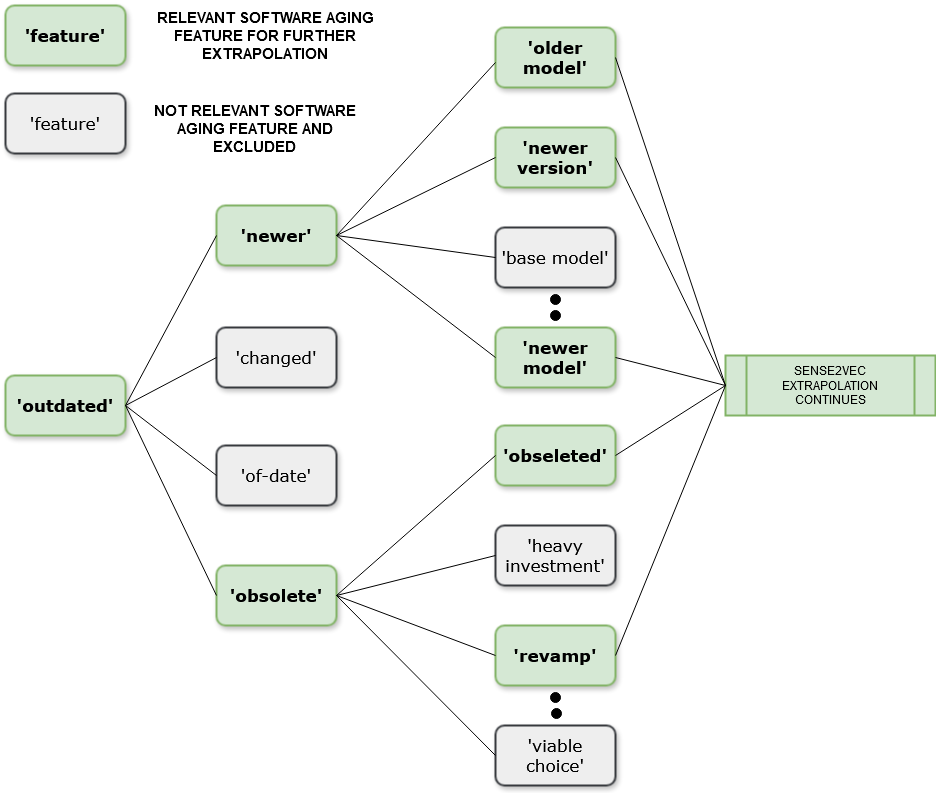}
  \caption{Sense2Vec: Feature Extrapolation Illustration}
  \label{sense2vec_feature_extrapolation}
\end{figure}
 
  We choose Sense2Vec due to its simple usage, and is trained on comments including software engineering from the Reddit platform. Sense2Vec is a variation of the popular Word2Vec model \citep{mikolov2013efficient}, which generates vector representations of words and phrases based on their contextual similarity. Unlike Word2Vec, Sense2Vec also considers Part-of-Speech tags to differentiate between words with multiple meanings (e.g., `code' as a noun versus a verb). In addition, Sense2Vec model is used in the academic literature for detecting TD within source code comments \citep{ sheikhaei2024empirical, sridharan2023pentacet}, making it particularly suited for our task. 
 
  %\end{enumerate}
  %\begin{enumerate}[label=\arabic{enumi}.\arabic{enumii}, start=2]
  %\setcounter{enumii}{2}
  \item \textbf{Manual Feature Verification:} During extrapolation, the Sense2Vec model may return features unrelated to software aging. To address this, we define criteria for inclusion in the expanded feature set based on how explicit the features are related to software aging, direct or indirect. Direct software aging features explicitly denote that software is becoming old or outdated. Examples include terms like `legacy', `obsolete', or `abandoned', indicating software has become old, outdated, no longer maintained, or no longer meets current performance, security, or compatibility requirements. Indirect software aging features indicate software aging implicitly, i.e., they are potential software aging-related features. Features such as `new program', `phasing', or `axed' are potential software aging-related features that need more contextual analysis but are not as explicit as direct software aging features. 
  %Similarly, features like `incompatibility' or `unusable' indicate difficulties operating with newer modules. Both direct and indirect terms reveal a need for updates or a comprehensive overhaul as the software's effectiveness declines over time.  

  \hspace*{2em} Although it can be challenging to determine whether indirect features truly indicate software aging, for example, `retooling' might result from software aging or simply indicating the code functionality, this ambiguity can only be resolved by examining the full comment text and, in some cases, the associated code context. Nevertheless, excluding such features too early may eliminate potentially valid indicators. Therefore, we include these features in the expanded set, understanding that they may be filtered out at later stages as we refine the AD features. 

  \hspace*{2em} The Algorithm \ref{alg:feature_extrapolation} lists the steps through which we obtain the manually verified software aging-related features. We start with an initial set of software aging features, i.e., the 8 seed features identified in step 1. For each seed feature, we find the top 30 software aging-related words. Each of these words is then manually verified by the first author to determine if it is directly or indirectly related to software aging. If a word qualifies, it is added to an `Expanded Features' set and also reintroduced into the seed list for further exploration. This process continues iteratively until there are no new software aging-related features identified, i.e., all features in the seed list have been utilized for extrapolation. This process expands the seed set by including new terms with each iteration.
  
  \begin{small}

\begin{minipage}{0.95\textwidth}
\begin{algorithm}[H]
\caption{Iterative Feature Extrapolation}
\label{alg:feature_extrapolation}
\begin{algorithmic}[1]
\REQUIRE Seed\_Features: Set of initial seed features related to software aging.
\ENSURE Expanded\_Features: Set of extrapolated features relevant to software aging.

\STATE Initialize Expanded\_Features $\gets$ \{\}
\WHILE{Seed\_Features is not empty}
    \FOR{feature $\in$ Seed\_Features}
        \STATE Input feature into Sense2Vec model.
        \STATE Retrieve top 30 Related\_Words $\gets$ Sense2Vec output.
        \FOR{word $\in$ Related\_Words}
            \STATE Manually verify if word is a direct or indirect software aging feature.
            \IF{word qualifies as direct or indirect software aging feature}
                \STATE Add word to Expanded\_Features.
                \STATE Add word to Seed\_Features for further extrapolation.
            \ELSE
                \STATE Exclude word from further consideration.
            \ENDIF
        \ENDFOR
        \STATE Remove feature from Seed\_Features (this feature has been processed).
    \ENDFOR
\ENDWHILE
\RETURN Expanded\_Features
\end{algorithmic}
\end{algorithm}
\end{minipage}
\end{small}

   Once all features have been processed, the `Expanded Features' set now includes both the initial 8 seed features and the newly discovered features associated with software aging. Overall, the first author identifies 151 software aging-related features through the iterative feature extrapolation algorithm \ref{alg:feature_extrapolation}. During the multi-author consensus of the identified 151 features, the second author identifies `malfunction', `crippled', `nonfunctional', `hot fix', `hotfix', and `inoperable' as not related to software aging. Finally, we identified 145 software aging-related features through this iterative feature extrapolation, of which 46 are direct and 99 are indirect software aging-related features. The complete list is included in the replication package \citep{muralisaad2025} and in Appendix \ref{sec:appendix_a}.  

\end{enumerate}

\subsubsection{B. Code Context Evaluated SAAD patterns:}\label{itm:b2}  
This sub-section explains Figure \ref{exploring_aad} part B. While the manually verified software aging features offer a starting point for detecting software aging-related comments, they do not serve as AD indicators. We query the PENTACET corpus using these features and browse the matched comments, including their surrounding code context to evaluate them for AD and then extract SAAD patterns from manually annotated SAAD comments. The objective of this step is to extract code context evaluated SAAD patterns from these manually annotated SAAD comments. The granular tasks in this step are elaborated in detail below. 
%\begin{enumerate}[label=\arabic{enumi},start=3]

\begin{enumerate}[label=\arabic*.]
\item First, we use the manually verified software aging-related extrapolated features from Sense2Vec in the previous step (including both direct and indirect aging-related features) to extract the natural language comments in the entire PENTACET dataset in an automated manner using a simple pattern matching technique.

\hspace*{2em} The pattern matching resulted in 993,016 matching comments out of 16,153,942 natural language comments. Table \ref{tab:matching_feature_annotation_feat}  shows the total counts of matching source code comments and features. Table \ref{tab:non_matching_feature_annotation_feat}  shows the list of features that did not return any matching comments in the PENTACET corpus. Table \ref{tab:manual_analysis_split} presents the distribution of features along with the count of their matching comments from the PENTACET corpus, separated into quartiles according to their frequency. 
%that were manually assessed for AD. 

\begin{table}[htbp!]
\caption{Extrapolated Software Aging Feature-Based Annotation: Matching Features}
  \label{tab:matching_feature_annotation_feat}
\centering
{\tiny
\renewcommand{\arraystretch}{1.5}
\begin{tabular}{cccc}
\hline
%\makecell{\textbf{Data} \\\textbf{Clusters}} & 
\makecell{\textbf{Data} } & 
\makecell{\textbf{\centering \# of Features} \\ \textbf{present} 
%\\ \textbf{in Data}
} & 
\makecell{\textbf{\centering \# of Comments} \\ \textbf{matching features} \\ 
%\textbf{in Data}
} & 
\makecell{\textbf{\centering \# of Features} \\ \textbf{not present} \\ 
%\textbf{in Data}
} \\ \hline
%\makecell{\textbf{Total Matching} \\ \textbf{Features}} & 
%\makecell{\textbf{Matching \# of} \\ \textbf{Comments} \\ \textbf{per Feature}} \\ \hline
\textbf{PENTACET} & \textbf{125} & \textbf{993,016} & \textbf{20} \\ \hline
\end{tabular}}
\end{table}

\begin{table}[htbp!]
\caption{Extrapolated Software Aging Feature-Based Annotation: Non-Matching Features}
  \label{tab:non_matching_feature_annotation_feat}
\centering
{\tiny
\renewcommand{\arraystretch}{1.5}
\begin{tabular}{cp{6cm}}
\hline
\makecell{\textbf{Data} } & 
\makecell{\textbf{\centering Non-Matching Software Aging } \\ \textbf{related features} 
%\\ \textbf{in Data}
} 

\\ \hline
\textbf{PENTACET} & \textbf{`newest patch', `modernise', `past release', `modernization', `outdated software', `antiquated', `security updates', `modernisation', `OS updates', `eradication', `overhauling', `dead-code', `nixed', `outdated system', `OS upgrades', `upgradability', `legacy software', `iOS updates', `end-of-life', `eradicating' } \\ \hline
\end{tabular}}
\end{table}

\begin{table}[htbp!]
\caption{Extrapolated Software Aging Features in OSS Comments (PENTACET corpus): Stats}
  \label{tab:manual_analysis_split}
\centering
{\tiny
\setlength{\tabcolsep}{2pt}
\begin{tabular}{cp{4.5cm}cc}
\hline
\makecell{\textbf{Software Aging} \\ \textbf{Features}} & 
\makecell{\textbf{\centering Matching Software Aging} \\ \textbf{related Features per quartile}} & 
\makecell{\textbf{Total \# of } \\ \textbf{Matching Features}} & 
\makecell{\textbf{Total \# of  } \\ \textbf{ Matching Comments} \\ \textbf{across features} \\ \textbf{per quartile} 
%\\ \textbf{manual anaysis}
} \\ \hline
%C1 & 
\multirow{8}{*}{\centering \makecell{\textbf{Quartile 1 (25\%)} \\ \textbf{Features with} \\ \textbf{ upto 5 comments}}}&
`ditched': 5, `old programs': 5, `upgraded version': 5, `unmaintained': 4, `eradicate': 4, `old software': 4, `newer system': 3, `latest model': 3, `security patches': 2, `upcoming version': 2, `Android updates': 2, `software upgrades': 2, `retooling': 1, `deserted': 1, `newest model': 1, `newer software': 1, `older software': 1, `retooled': 1, `nixing': 1
& \multirow{8}{*}{\centering 39} & \multirow{8}{*}{\centering 48} \\ \hline
\multirow{16}{*}{\centering\makecell{\textbf{Quartile 2 (50\%)} \\ \textbf{5$>$Features$<=$83} }} &
`phasing': 83, `new program': 83, `new platform': 75, `retrofitted': 70, `legacy system': 69, `decommission': 61, `legacy applications': 52, `overhaul': 46, `legacy apps': 45, `abandoning': 45, `next major release': 45, `previous model': 42, `old system': 40, `retool': 36, `neglected': 35, `newer model': 28, `archaic': 26, `trashed': 24, `past version': 24, `scrapped': 22, `modernize': 21, `decommissioned': 20, `older model': 17, `latest build': 17, `improved version': 16, `OTA updates': 15, `latest patch': 14, `ditching': 10, `outdated version': 10, `legacy systems': 9, `older system': 8, `new software': 8, `next revision': 7, `software updates': 6, `newest update': 6
 & \multirow{16}{*}{\centering 35} & \multirow{16}{*}{\centering 1,135} \\ \hline
\multirow{16}{*}{\centering \makecell{\textbf{Quartile 3 (75\%)} \\ \textbf{83$>$Features$<=$690} }} & `newer version': 690, `outdated': 687, `trash': 672, `new feature': 653, `future versions': 621, `abandoned': 529, `earlier version': 484, `new system': 450, `unusable': 441, `retrofit': 405, `old code': 351, `eliminating': 328, `new implementation': 287, `later version': 281, `prior version': 247, `upgradable': 241, `phased': 221, `ditch': 210, `obsoleted': 208, `legacy code': 207, `purging': 205, `redesign': 180, `next version': 160, `compatible version': 156, `legacy support': 150, `old model': 126, `updated version': 126, `next release': 125, `shelved': 116, `axing': 114, `latest update': 107, `deadcode': 97, `newest version': 93, `firmware update': 90, `revamp': 89
 & \multirow{16}{*}{\centering 35} & \multirow{16}{*}{\centering 10,147} \\ \hline
\multirow{16}{*}{\centering \makecell{\textbf{Quartile 4 ($>$75\%)} \\ \textbf{Features with} \\ \textbf{ greater than } \\ \textbf{690 comments}}} & 'update': 194722, 'old': 183431, 'remove': 172514, 'rid': 159009, 'delete': 75633, 'updated': 42917, 'patch': 29774, 'deprecated': 21866, 'updating': 10801, 'removing': 10524, 'legacy': 9435, 'unused': 8603, 'upgrade': 8151, 'eliminate': 5261, 'purge': 4977, 'deleting': 4891, 'unnecessary': 4338, 'rebuild':3959, 'newer': 3512, 'backward compatibility': 3351, 'obsolete': 3069, 'eol': 2785, 'backwards compatibility': 2196, 'new version': 2018, 'useless': 2012, 'future release': 1946, 'latest version': 1480, 'previous version': 1307, 'old version': 1093, 'scrap': 1085, 'future version': 1060, 'older version': 929, 'axed': 873, 'upgrading': 733, 'unneeded': 729, 'new functionality': 702
& \multirow{16}{*}{\centering 36} & \multirow{16}{*}{\centering 981,686} \\ \hline
\textbf{TOTAL} & & \textbf{125} & \textbf{993,016} \\ \hline
\end{tabular}
}
\end{table}

Although the software aging-related features successfully identified a large number of matching comments, the lack of a complete understanding of the surrounding code context means these results are raw and unfiltered. If such comments are categorized as SAAD comments without further evaluation of the surrounding code context, it poses a significant risk of including a lot of false positives. %There may be many comments that contain keywords related to software aging, but are not genuinely indicative of AD unless the entire comment is fully evaluated. 
A few illustrations explaining the need for the analysis of source code comments along with the code context, 

%In some cases, evaluating the comment text alone may not be sufficient and requires an understanding of the surrounding code context to make an accurate determination. 

%The need to investigate the full comment text along with the source code context is crucial for identifying real SAAD comments. However
%\vspace{-0.5cm}

%\begin{table}[htbp!]
%\caption{Extrapolated Software Aging Features-Based Annotation: Matching Comments}
  %\label{tab:raw_feature_annotation_comments}
%\centering
%{\tiny
%\renewcommand{\arraystretch}{1.5}
%\begin{tabular}{cc}
%\hline
%\makecell{\textbf{PENTACET} \\\textbf{Cluster}} & 

%\makecell{\textbf{\centering \# of Comments} \\ \textbf{matching features} \\ \textbf{in Data}} \\ \hline

%\textbf{1} &  \textbf{50,375}  \\
%\textbf{2} &  \textbf{100,713}  \\ 
%\textbf{3} &  \textbf{203,607}  \\ 
%\textbf{4} &  \textbf{638,021}  \\ \hline
%\textbf{Total} &  \textbf{992,716}  \\ 
%\hline
%\end{tabular}}
%\end{table}

%\vspace{-1cm}

\begin{enumerate}[label=\alph*.,]
\item Feature `unused' along with its matching full text comment and its associated source code,
\begin{lstlisting}
while(jumpSite.getPrevious() != null &&
jumpSite.getPrevious() != gotoNode)
clinit.instructions.remove(jumpSite.getPrevious());
// Remove jumpsite if unused
boolean used = false;
\end{lstlisting}

The feature `unused' by itself seems to be related to software aging (indicating unused code, i.e., the code once functional has become irrelevant but still exists in the source code) but the matched full-text comment is actually an explanatory note on the code logic. Analyzing along with the source code context clearly indicates that the comment is simply an explanatory comment but does not indicate AD.

\item Another example feature `useless' along with its matching full text comment and its associated source code,

\begin{lstlisting}[basicstyle=\small\ttfamily]
if (!uri.endsWith("".jar"")) {
// the maven plugin is putting some useless source url sometimes... 
log(Message.MSG_WARN, ""A source uri is suspect, it is not ending 
with .jar, it is probably"" + "" a pointer to a download page. 
Ignoring it."");
return;
}
\end{lstlisting}

The feature `useless', although it seems software aging-related, the comment text along with code context indicates it as an explanatory comment explaining conditional logic.

\end{enumerate}

Therefore, to reduce such false positives, it is essential to identify code context evaluated SAAD patterns after thoroughly analyzing the comments along with their surrounding code context, which will reduce false positives considerably and reveal genuine SAAD comments.

\item 

%an iterative cycle in which a random sample of comments is selected without repetition. 
 
The next step is to manually annotate the candidate SAAD comments (i.e., those that matched the automated pattern matching using manually verified software aging features from step A) to extract SAAD patterns. If a manually annotated comment, along with its corresponding source code context, is found to be AD, a text pattern is extracted from the comment that reflects the core theme of the associated AD. This step constitutes the inductive open coding \citep{seaman1999qualitative, williams2019art} that assigns codes (patterns) emerging from the data (source code comments). Below are a few SAAD comments with the extracted SAAD patterns to illustrate SAAD pattern (open coding) extraction during manual annotation.
\begin{itemize}
    \item \begin{lstlisting}[basicstyle=\small\ttfamily]
    // Keep this for legacy code.
    \end{lstlisting}
    Extracted SAAD pattern: `for legacy'

    \item \begin{lstlisting}[basicstyle=\small\ttfamily]
    // Obsoleted method which does nothing.
    \end{lstlisting}
    Extracted SAAD pattern: 
    `\texttt{\^{}}obsolete' (i.e., comments that begin with obsolete.)

    \item \begin{lstlisting}[basicstyle=\small\ttfamily]
    // couldn't release lock. No problem, this is legacy 
    // code anyways.
    \end{lstlisting}
    Extracted SAAD pattern: `is legacy'
\end{itemize}    
Table \ref{tab:pattern_extraction} lists the breakdown of the SAAD pattern extraction after manual analysis of source code comments, along with the source code context. Finally, this manually verified SAAD pattern extraction step results in a total of 399 SAAD patterns and 1,708 gold standard SAAD comments.

\end{enumerate}

\subsubsection{C. Code Context Evaluated SAAD Comments:} \label{itm:3}
This subsection explains Figure \ref{exploring_aad} part C. 
The objective of this step is to utilize the SAAD patterns from step \ref{itm:b2} to perform automated heuristics-based (pattern matching) SAAD detection and validate the matching comments, resulting in gold and silver standard SAAD comments as explained by \citep{wissler2014gold}. Gold standard SAAD comments are fully manually verified, while silver standard SAAD comments are machine-predicted and validated through iterative sampling until a consistent quality criterion is met.
\begin{enumerate}[label=\arabic*.]
  \item \textbf{Automated SAAD pattern Matching:} We utilize the 399 code context evaluated SAAD patterns to annotate the natural language code comments in the entire PENTACET corpus using the simple pattern matching technique. 
  \item \textbf{Manual Verification of SAAD comments:} Although the utilized SAAD patterns in pattern matching are manually extracted after code context evaluation, heuristics-based annotation is not foolproof and may include false positives that warrant manual verification. For manual verification of the candidate SAAD comments (i.e., matched comments using SAAD patterns), the first author takes a non-repeating random sample stratified by the SAAD patterns of the candidate SAAD comments (to ensure the sample reflects the same distribution of the entire matched SAAD comments) and evaluates if the comment is truly a SAAD comment or not (NON-SAAD) after evaluating the code context. The sample size is determined at 95\% confidence level with a 5\% margin of error, using

\[ n = \left( \frac{z^2 \cdot p \cdot (1 - p)}{E^2} \right) \tag{I} \]

where:
\begin{itemize}
  \item \( n \) is the required sample size
  \item \( z \) is the z-score corresponding to the desired confidence level (1.96 for 95\% confidence)
  \item \( p \) is the estimated proportion of the population that is SAAD
  \item \( E \) is the margin of error (0.05 for a 5\% error rate)
\end{itemize}

By treating the manual annotation of pattern-based SAAD comments as ground truth labels, we calculate the F1 score for the predicted SAAD comments using SAAD patterns. We choose F1, the harmonic mean of precision and recall because it is insensitive to true negatives while focusing mainly on true positives, false positives, and false negatives \citep{canbek2017binary}.
\begin{itemize}
    \item If the F1 score for SAAD automated annotation using SAAD patterns is less than 95\%, we refine the SAAD patterns for automated annotation by excluding those that yield more than 25\% false positive (FP) SAAD comments and then repeat the automated annotation of the entire PENTACET corpus using the refined SAAD patterns. We choose a 25\% FP rate as the exclusion threshold for SAAD patterns, considering the trade-off between precision and recall in the automated detection of SAAD comments. A stricter filtering criteria, such as 15\% or 5\%, can exclude SAAD patterns that capture true positives i.e., actual SAAD comments, thereby reducing recall and limiting the comprehensiveness of SAAD patterns for future reuse. However, the overall F1 score remains above 95\% for overall pattern-based SAAD detection.
    \item We stop this iterative cycle of automated annotation using SAAD patterns and manual verification once the automated detection of SAAD comments using SAAD patterns consistently achieves an F1 score of at least 95\%. Table \ref{tab:pattern_analysis} lists the results of iterative cycle of SAAD detection. The manual annotations of heuristics-based SAAD detection results in 1,636 SAAD comments.
\end{itemize}    

\item \textbf{SAAD Corner Cases:} During the evaluation of heuristics-based SAAD detection (i.e., the manual annotation of SAAD comments for ground truth), the first author discussed corner cases with all the authors. %We did not reach consensus among all authors regarding scenarios that involve potential future design changes as AD. 
For example, consider the following comment, which is marked as SAAD for the pattern `future versions of the'. 

\begin{lstlisting}[basicstyle=\small\ttfamily]
/** Users can create a 
*mechanism for managing this, or relinquish the use of ""==""
*and use the .sameNodeAs() mechanism, which is under 
*consideration for future versions of the DOM.   */
\end{lstlisting}

Although the comment mainly focuses on a potential future design change, it is not an SAAD comment. All the authors agreed that comments that indicate potential design change for the future are explanatory comments but not SAAD comments. The first author revisited the gold standard manual annotations of SAAD comments for SAAD pattern extraction made in the previous step \ref{itm:b2} to exclude such scenarios as explanatory comments rather than SAAD. Overall, the combined, i.e., after correcting the identified corner cases and removing redundant comments from the manually verified SAAD comments from steps \ref{itm:b2} and \ref{itm:3} results in a total of 2,562 SAAD comments.

  \item %Once multiple iterations of manual labeling are completed and the F1-score consistently reaches greater than 0.90 after revising the incremental filtering criteria, 
  \textbf{SAAD Annotation Quality:} The data quality validation process involves all authors to ensure consensus and eliminate any personal bias. The quality of the manual annotations from steps \ref{itm:b2} (SAAD pattern extraction) and \ref{itm:3} (pattern-based SAAD comments detection) is assessed using the Inter-Rater Agreement (IRA) score, calculated with the Cohen's Kappa statistic, which yields a substantial value of 0.695 \citep{emam1999benchmarking}, thus assuring the quality of the data for further analysis in this empirical study. The disagreements from the initial analysis were subsequently resolved through discussions among all authors, leading to a consensus.

\end{enumerate}

\subsection{Results}
This section contains the results of the three steps involved in detecting SAAD comments, as discussed in Section \ref{sec:det_saad_approach}, using text patterns. It includes, A. Software Aging features, B. Code Context evaluated SAAD patterns and C. Code Context evaluated SAAD comments.
\paragraph{A. Software Aging Features}
\label{a_aging_features}
It is observed from Table \ref{tab:matching_feature_annotation_feat} that out of 145 features, 125 returned matching comments and 20 did not return any matching comments. Although these 20 features listed in Table \ref{tab:non_matching_feature_annotation_feat} did not return any matching comment, they are related to software aging; it is possible that there does not exist any comment with specific combination words such as `newest patch' or `legacy software', etc.,

Table \ref{tab:manual_analysis_split} shows that the frequency of the software aging features in quartile 1 is very low with 5 matching comments, indicating they rarely occur. The features in quartiles 2 and 3 occur moderately in the PENTACET corpus, with matching comments ranging from 5 up to 690. The quartile 4 lists the most frequently occurring software aging features with over 690 matching comments each. Overall, the 125 features results in identifying 993,016 software aging-related source code comments.

\paragraph{B. Code Context Evaluated SAAD patterns}
\label{b_ad_patterns}
Table \ref{tab:pattern_extraction} lists the manually verified software aging-related feature matching comments and the corresponding count of SAAD patterns extracted from comments grouped under each quartile from Table \ref{tab:manual_analysis_split}. The aging feature matching comments from quartiles 1 (48) and 2 (1135) are relatively fewer compared to comments from quartiles 3 (10,147) and 4 (981,686) indicating the matching software aging features occur rarely. The first author manually annotated all the source code comments from quartiles 1 and 2 for AD since they are relatively lesser in numbers. The manual annotation resulted in 15 and 254 SAAD comments from 11 and 92 SAAD patterns for quartiles 1 and 2, respectively. %The manual annotation identified 11 and 92 SAAD patterns for quartiles 1 and 2, respectively. 

\begin{table}[htbp]
\caption{Code Context Evaluated SAAD pattern Extraction: Stats}
\label{tab:pattern_extraction}
\resizebox{\columnwidth}{!}{%
\begin{tabular}{c|c|c|c|c|c}
\hline
\textbf{\begin{tabular}[c]{@{}c@{}}Quartile \#\\ (Total \# \\ Comments)\end{tabular}} &
  \textbf{\begin{tabular}[c]{@{}c@{}}All Comments/\\ Stratified \\ Random \\ Sample \#\end{tabular}} &
  \textbf{\begin{tabular}[c]{@{}c@{}}Total \\ Comments\\ Manually \\ Annotated\end{tabular}} &
  \textbf{\begin{tabular}[c]{@{}c@{}}Total AD\\ Comments\end{tabular}} &
  \textbf{\begin{tabular}[c]{@{}c@{}}New AD\\ Patterns\\ Emerged\end{tabular}} &
  \textbf{\begin{tabular}[c]{@{}c@{}}\% New AD\\ Patterns\end{tabular}} \\ \hline
\textbf{\begin{tabular}[c]{@{}c@{}}Q1\\ (48)\end{tabular}}                        & All Comments & 48        & 15            & 11           & 73.33 \\ \hline
\textbf{\begin{tabular}[c]{@{}c@{}}Q2\\ (1135)\end{tabular}}                      & All Comments & 1135      & 254           & 92           & 36.22 \\ \hline
\multirow{10}{*}{\textbf{\begin{tabular}[c]{@{}c@{}}Q3\\ (10,147)\end{tabular}}}  & Sample 1     & 385       & 76            & 52           & 68.42 \\
                                                                                  & Sample 2     & 385       & 76            & 33           & 43.42 \\
                                                                                  & Sample 3     & 385       & 75            & 30           & 40.00 \\
                                                                                  & Sample 4     & 385       & 63            & 25           & 39.68 \\
                                                                                  & Sample 5     & 385       & 74            & 17           & 22.97 \\
                                                                                  & Sample 6     & 385       & 58            & 11           & 18.97 \\
                                                                                  & Sample 7     & 385       & 51            & 10           & 19.61 \\
                                                                                  & Sample 8     & 385       & 50            & 5            & 10.00 \\
                                                                                  & Sample 9     & 385       & 48            & 3            & 6.25  \\
                                                                                  & Sample 10    & 385       & 56            & 2            & 3.57  \\ \hline
\multirow{11}{*}{\textbf{\begin{tabular}[c]{@{}c@{}}Q4\\ (981,686)\end{tabular}}} & Sample 1     & 385       & 75            & 51           & 68.00 \\
                                                                                  & Sample 2     & 385       & 89            & 31           & 34.83 \\
                                                                                  & Sample 3     & 385       & 90            & 24           & 26.67 \\
                                                                                  & Sample 4     & 385       & 86            & 17           & 19.77 \\
                                                                                  & Sample 5     & 385       & 66            & 12           & 18.18 \\
                                                                                  & Sample 6     & 385       & 80            & 11           & 13.75 \\
                                                                                  & Sample 7     & 385       & 66            & 9            & 13.64 \\
                                                                                  & Sample 8     & 385       & 66            & 9            & 13.64 \\
                                                                                  & Sample 9     & 385       & 63            & 7            & 11.11 \\
                                                                                  & Sample 10    & 385       & 68            & 8            & 11.76 \\
                                                                                  & Sample 11    & 385       & 63            & 3            & 4.76  \\ \hline
\textbf{Total}                                                                    & \textbf{}    & \textbf{} & \textbf{1708} & \textbf{473} &       \\ 
\end{tabular}%
}
\end{table}
For quartiles 3 and 4, the manual annotation of all comments is enormous and practically infeasible, as it would require manually analyzing 991,833 comments along with their code context. Therefore, we adopt an iterative sampling and manual annotation approach to evaluate the comments in the sample for AD. 
The first author takes a random sample stratified by the frequency of the matching software aging features as listed in Table \ref{tab:manual_analysis_split} to ensure that the sample represents the same distribution of the matching comments in the quartile. The sample size is determined at 95\% confidence level with a 5\% margin of error, using equation I, as 385.

The first author repeats iterative sampling and manual analysis until no more than 5\% new SAAD patterns emerge from a sample. This threshold as a stopping criterion ensures that most SAAD patterns are identified without missing significant ones. The iterative sampling and manual annotation initially identified 71.05\% and 68\% new SAAD patterns for quartiles 3 and 4, respectively; however, these percentages steadily declined with each iteration, reaching the stopping criterion at the 10th sample for quartile 3 (3.57\% new SAAD patterns) and at the 11th sample for quartile 4 (4.76\% new SAAD patterns). Although our iterative sampling approach differs from analyzing all 991,813 comments in quartiles 3 and 4, it is more efficient given the time and resources required for the manual analysis of such a large dataset. Furthermore, using a stopping criterion of fewer than 5\% new SAAD patterns ensures that significant SAAD patterns are already captured through iterative sampling and subsequent manual annotations.

The manual verification of comments with source code context initially results in the extraction of 473 SAAD patterns and 1,708 manually verified SAAD comments. Then, removing duplicate SAAD patterns and simplifying text patterns, for example, merging `for older version' and `for older versions' into one pattern, `for older versions?' where the `?' indicates that the character `s' is optional) reduces the number of code context-evaluated SAAD patterns to 399.

\paragraph{C. Code Context Evaluated SAAD Comments}
\label{c_saad_comments}

Tables \ref{tab:pattern_analysis} contain the validation metrics (obtained from the manual annotation of automated SAAD predictions with code context, that serves as ground truth). We utilize precision, recall and F1 as described below for our validation of detected SAAD comments,
%and the confusion matrix for the last iteration's F1 score, that consistently achieved at least 0.95 F1 (as highlighted in bold in Table \ref{tab:pattern_analysis})

\textbf{PRECISION:} This indicates the number of items identified as positive (in this case, comments indicating AD) are actually positive.

\textbf{RECALL:} This represents how many positive cases were identified correctly.

\textbf{F1:} This is the harmonic mean of precision and recall. It gives a balanced measure of the automated annotation's accuracy.

In the first iteration utilizing 399 SAAD patterns for the heuristics-based SAAD detection, we achieved a precision of 0.803, a recall of 1.00, and an F1 score of 0.891 on 44,181 SAAD comments. The recall remains at 1.00, indicating that only comments that match the SAAD patterns are included in the sample for manual evaluation. This ensures that manual evaluation focuses on these matched comments, which are expected to consistently achieve an F1 score of at least 0.95 in the SAAD pattern-based SAAD detection. We exclude any SAAD patterns that result in at least 25\% false positives. After excluding 22 such SAAD patterns in the first iteration, the pattern-based SAAD detection is re-initiated using 377 SAAD patterns for the entire PENTACET corpus.

\begin{table}[htbp]
\caption{Heuristics-Based SAAD Detection: SAAD Pattern Analysis}
\label{tab:pattern_analysis}
\resizebox{\columnwidth}{!}{%
\begin{tabular}{c|c|c|c|c|l}
\hline
\textbf{Iteration} &
  \textbf{Precision} &
  \textbf{Recall} &
  \textbf{F1} &
  \textbf{\#SAAD comments} &
  \multicolumn{1}{c}{\textbf{\begin{tabular}[c]{@{}c@{}}Excluded\\ SAAD patterns\end{tabular}}} \\ \hline
1 &
  0.795 &
  1.00 &
  0.886 &
  44,181 &
  \begin{tabular}[c]{@{}l@{}}changes? in future, currently doesn't support, in newer versions, \\ cannot be used in, are deprecated, consider updating, this is not used, \\ upgrade to the latest, not needed anymore, old versions? of, is obsolete, \\ future versions of the, may be deprecated, previous versions? of, \\ needs? to be updated, by older versions?, old behavior, \textasciicircum{}old version,\\  older versions? of, use the old, for older versions?, or older version\end{tabular} \\ \hline
2 &
  0.803 &
  1.00 &
  0.891 &
  38,619 &
  receive a major overhaul, for obsolete, old code that, the old code \\ \hline
3 &
  0.795 &
  1.00 &
  0.886 &
  38,454 &
  \begin{tabular}[c]{@{}l@{}}with future versions of, will be added in future, \textasciicircum{}old unsupported version,\\ deprecated on, on old(er)? versions?, 'obsoleted, but not yet removed',  \\ an earlier version of, due to be decommissioned, future versions of sdc, \\ has been retrofitted, being phased out, in{[}.\textbackslash{}s\textbackslash{}w{]}*future versions? of,\\ compatible with future, on older systems,  revamp the, is legacy,\\ note{[}\textbackslash{}s\textbackslash{}w:{]}*legacy, available in a future,  \textasciicircum{}outdated, is unnecessary\$,  \\ from older version, legacy code that\end{tabular} \\ \hline
4 &
  0.935 &
  1.00 &
  0.966 &
  37,234 &
  \begin{tabular}[c]{@{}l@{}}upgrade from, supported in a future, not really needed, \\ currently not supported,  :{[}.\textbackslash{}s\textbackslash{}w{]}*remove{[}.{]}*\$, remove unnecessary, \\ \textasciicircum{}obsolete, this{[}\textbackslash{}s\textbackslash{}w{]}*will be removed\textbackslash{}s*(in$|$after$|$\textbackslash{}.)?\end{tabular} \\ \hline
5 &
  0.927 &
  1.00 &
  \textbf{0.962} &
  35,630 &
  \multicolumn{1}{c}{\textbf{\begin{tabular}[c]{@{}c@{}}CONSISTENT F1 \textgreater{}0.95 ACHIEVED \\ FOR HEURISTICS-BASED SAAD DETECTION\end{tabular}}} \\ \hline
\end{tabular}%
}
\end{table}

The iterative refinement of SAAD patterns identified more false positives, reflecting that the total number of SAAD comments gradually decreased after each iteration while the F1 score improved. By the fourth iteration, precision increased to 0.935, and the F1 score reached 0.966, indicating that nearly all SAAD comments were captured with fewer false positives. In the final (fifth) iteration, despite a small drop in precision, the F1 score of 0.962 remains consistently above 0.95, indicating that the detection performance has converged on 35,630 SAAD comments.

\subsection{Discussion}
\hspace*{2em} We developed a novel approach to identify SAAD by analyzing text features in source code comments. We studied software aging explicitly expressed in source code comments left by developers after evaluating the code context of the comment to determine SAAD (i.e., AD expressed in code comments). Our method achieved an F1 score of 0.962, which is encouraging. However, this result should be interpreted with caution due to potential false positives; some code comments categorized as SAAD may not actually indicate SAAD but merely serve as explanatory comments explaining the code logic. Currently, our approach primarily utilizes basic pattern recognition techniques. We adopt this simple method because our work is the first exploratory effort to mine software aging features (145 features) and code context evaluated SAAD patterns (399) from source code comments for detecting SAAD comments. As part of this effort, we compiled a gold standard SAAD dataset (comprising 2,562 SAAD comments) which is fully manually verified, and a silver standard SAAD dataset (comprising 35,630 SAAD comments) evaluated through iterative sampling and subsequent manual verification of the sampled comments until consistent detection performance was achieved. For this initial effort, relying on pattern recognition to detect SAAD in comments has proven sufficient. 

\hspace*{2em}Future research could leverage the gold standard SAAD dataset to employ more sophisticated methods, such as advanced deep learning techniques, to further enhance the accuracy and effectiveness of SAAD detection. We utilize the 399 SAAD patterns and the silver standard SAAD dataset comprising 35,630 SAAD comments along with their context in our subsequent research questions to classify the SAAD into types and quantify the detected SAAD types.
\begin{tcolorbox}
\textbf{Implications:} Analysis of natural language comments in source code offers a unique way to detect software aging and its related debt, revealing areas that may need refactoring or removal. This approach offers software practitioners valuable insights into software health. For researchers, future research can focus on more effective ways of detecting SAAD across multiple programming languages.
\end{tcolorbox}

\section{RQ2: SAAD Classification}

\label{aging_taxonomy}

\subsection{Motivation}
RQ2: How can SAAD in source code comments be classified?

\hspace*{2em} Once AD is detected in source code comments, the next logical step is understanding its nature and implications. The nature of AD in the source code comments varies from comments indicating deprecated methods to unused or obsolete code. Each type of aging offers varying insights into software health. %Therefore, the classification of AD is essential for assessing its severity and for the required remediation efforts.
After detecting SAAD comments, merely grouping them under the category SAAD is inefficient since different types of AD may require different efforts. For instance, rectifying an obsolete library call necessitates more detailed steps than managing legacy code. In addition, the categorization of the identified AD in SAAD comments offers leverage to software developers to effectively prioritize the remediation efforts. Therefore, we introduce a taxonomy for SAAD that offers a clear structure and a practical framework for managing and mitigating the impacts of different types of software aging.

\subsection{Approach}
\begin{figure}[h]
    \centering
    \includegraphics[width=0.65\linewidth]{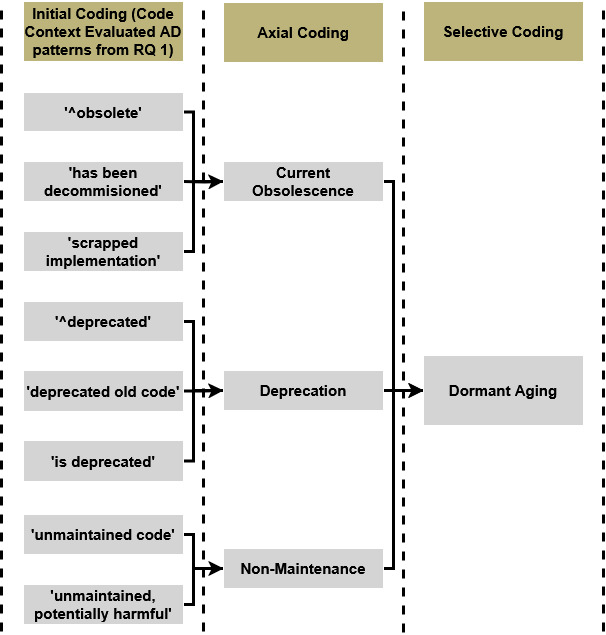}
    \caption{Methodology: Bottom-up coding used to develop the taxonomy}
    \label{fig:taxonomy_coding}
\end{figure}

To define a robust taxonomy of AD grounded in actual source code comments, we adopt qualitative analysis techniques and build a bottom-up classification starting from the validated aging features identified in RQ 1. Figure \ref{fig:taxonomy_coding} explains the methodology used to develop the AD taxonomy. We begin by adopting a deductive approach to coding \citep{miles1994qualitative}, using preliminary codes from Section \ref{b_ad_patterns} (i.e., 399 code context-evaluated SAAD patterns). Once these preliminary codes are identified, we delve deeper to discover the relationships among the categories through axial coding \citep{williams2019art}. The first author iteratively examines each of the 399 SAAD patterns and groups them into emerging categories, identified through axial coding, which allows us to understand the broader structures present in the data. The second author then reviews the arrangement of SAAD patterns in their respective categories to mitigate any potential bias in grouping. Once the categories are established, we implement selective coding \citep{williams2019art} to refine and finalize the core categories. Our focus is to ensure that the resulting taxonomy is both comprehensive and cohesive.  %We revised the taxonomy through multiple iterations utilizing a Miro board\footnote{\url{https://miro.com/app/board/uXjVNXE5RSg=/?share_link_id=34711778669}}. 

\subsection{Results}
\begin{figure*}
    \centering
    \includegraphics[width=0.95\textwidth]{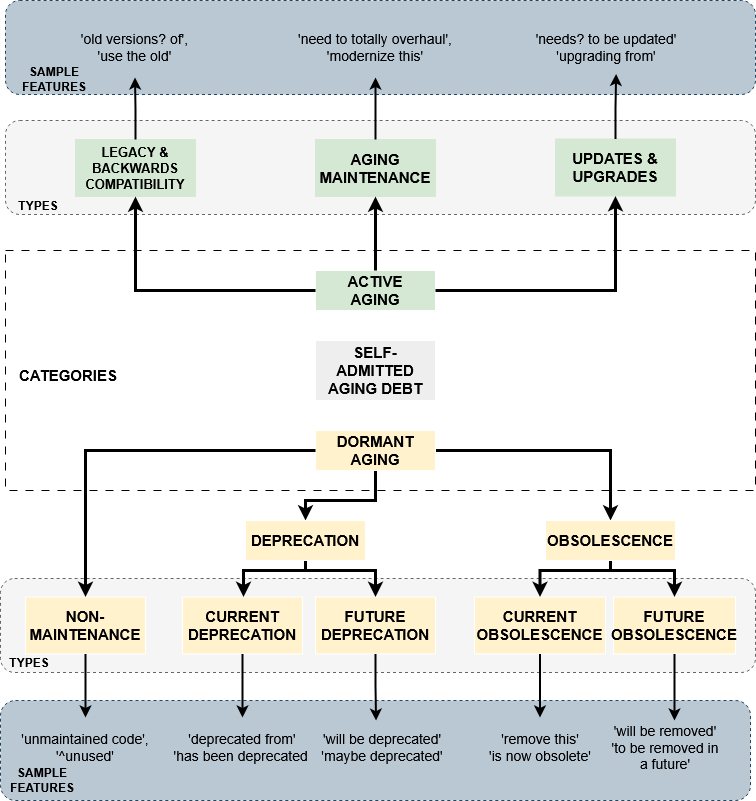}
    \caption{Aging Taxonomy}
    \label{fig:taxonomy}
\end{figure*}

Figure \ref{fig:taxonomy} shows the developed taxonomy of AD with its categories and types developed using the code context evaluated 399 SAAD patterns from RQ 1. Our proposed SAAD taxonomy is structured under two categories `Active' and `Dormancy' based on the code context evaluated SAAD patterns. 

\subsubsection{Active Aging}
Active aging refers to those segments of software that remain functional but exhibit signs of aging. %It includes refactoring need for legacy and outdated code.
It includes ongoing refactoring to support legacy or old systems, continuous modernization to accommodate dependency or platform updates. This category reflects active development efforts to keep the software current and operational.

\textbf{Updates \& Upgrades AD}: This refers to the debt incurred when modules and components must be updated or upgraded due to evolving dependencies. Typically, this debt highlights those sections of software that are deliberately in need of updates (refers to minor patches) and upgrades (refers to major functionality changes or enhancements). These modifications ensure that the software is aligned with evolving technological standards and dependencies. %Unlike Version lag \cite{zerouali2018empirical}, which focuses on version log or history to upgrade to the latest available version of the software package, Updates and Upgrades SAAD utilizes the source code comments left by the developers in the source code to detect Updates \& Upgrades SAAD. Utilizing source code comments offers captures developers insights can highlight specific component or functionality and indicate the need for updates or upgrades before a version change occurs which will not be in version history.
    Examples:
    \begin{lstlisting}
//Must develop a strategy for upgrading from older 
// SubscriptionWrapper versions to newer versions.

// we'll probably have to remove 'diff 0' after upgrading to 
//lucene 3.1.
    \end{lstlisting}
\textbf{Aging Maintenance}: 
    Aging Maintenance AD refers to the debt incurred from refactoring, retrofitting, or modernizing the source code to stay current. It includes overhauling or adapting existing code to handle new changes or planned improvements for upcoming software versions. 
    
    %Unlike Legacy and Compatibility AD, which explicitly covers the effort required to maintain outdated legacy systems for compatibility, Aging Maintenance AD focuses on keeping aging (i.e., current but aging) code relevant. While Updates and Upgrades AD addresses the effort needed to remain current with external dependency-related code, and Past-Version Specific AD covers adjustments needed for features and functionalities internal to the software from past software versions, Aging Maintenance AD accounts for the ongoing upkeep of present code.
    %This involves managing aging software components, addressing those that are momentarily non-functional, lack support, require future attention, or are unnecessary for specific purposes. The goal is to ensure that the software remains adaptable and functional despite showing signs of aging. Examples:
    \begin{lstlisting}
//Some of these internal IDs are outdated and don't represent 
//what these challenges do.

//This is pretty old code and might need some upgrades.
    \end{lstlisting}
\textbf{Legacy \& Backwards Compatibility}: This addresses the need for current software systems to maintain interoperability with older systems or components. The main focus of this debt is to cover the efforts required to keep the outdated systems (legacy systems) that are still crucial to business are in sync with the current systems (technological landscape). %that as software evolves, it does not alienate existing set-ups or requires significant overhauls to retain compatibility. 
    Examples:
    \begin{lstlisting}
//As of Java 2 platform v 1.4, this class is now obsolete, 
//doesn't do anything, and is only included for backwards 
//API compatibility.

// For backward compatibility generate an empty paint event. Not 
// doing this broke parts of Netbeans.


    \end{lstlisting}

%\subsubsection{Legacy} Legacy software refers to older systems, applications, or technologies that are still in use even though they might be outdated or replaced by newer alternatives. Legacy software can be challenging to maintain, update, and integrate with modern systems.

%\subsubsection{Backwards Compatibility} Backwards compatibility refers to the ability of a software or system to function properly with earlier versions of itself or to support data and interfaces from previous iterations. This ensures that users can upgrade without losing access to existing data or functionality.

%\subsection{Dormant Aging} In the "dormancy" state, a software system is no longer actively maintained, developed, or used. It may be deprecated, obsolete, or unsupported. Software components in this state are not receiving updates or improvements, and they might not be compatible with newer technologies or systems. Dormancy can arise due to changes in business needs, technological advancements, or a shift in priorities.

\subsubsection{Dormant Aging} This refers to parts of the software that are either already obsolete or becoming inactive and eventually obsolete. Although such sections may not cause immediate issues, they represent a deferred maintenance burden in the form of refactoring or removal. We refer to the pending maintenance effort associated with these items as Dormant SAAD. 
%Certain software components age and become vulnerable without regular patches \cite{wang2019detecting}. The Open Web Application Security Project (OWASP) recently highlighted \textbf{A06:2021 Vulnerable and Outdated  Components} \footnote{\url{https://owasp.org/www-project-top-ten/}} as one among the top security risk for web apps, citing unawareness of software versions, outdated software, infrequent vulnerability scans, and patch updates. %which we term as Dormant SAAD.  This includes sections of code that are or may become deprecated, obsolete, unused, or unsupported. Such software components do not receive updates or improvements and may not be compatible with newer technologies or systems. Dormant AD typically refers to dead code or features that no longer play an active role in the system but remain in the source code, potentially causing clutter, or even harmful side effects~\cite{romano2018multi, caivano2021exploratory}. %Dormancy can arise due to changes in business needs, technological advancements, or shifts in priorities.

\textbf{Deprecation:}  %Deprecation refers to the process of indicating that a particular feature, component, or functionality is no longer recommended for use. When such deprecated features and functionalities are used, there exists a hidden cost in terms of efforts to replace them with current alternates. In addition, all deprecated features are tied with a maintenance overhead to remove their usage once their support ends. 
This refers to the debt associated with deprecated software components. The maintenance overhead of temporarily maintaining the deprecated component and the implied maintenance effort required to eventually remove it constitutes Deprecation AD. %Deprecation AD refers to the refactoring efforts caused by the deprecated features and functionalities that are still being used in the software.  %It serves as a warning to users and developers that the deprecated element should be avoided in favor of alternative solutions. 
Deprecation AD manifests in two different forms: Current Deprecation, referring to sections of code that are already deprecated, and Future Deprecation, referring to sections of code that are expected to be deprecated. Examples of future and current deprecations:
        \begin{lstlisting}
// this should be deprecated - jcs

// Object[] and List are outdated and may be deprecated some day


// everything below is deprecated

// deprecated method

    \end{lstlisting}

\textbf{Obsolescence:} This refers to the debt associated with refactoring code that results from the removal of obsolete sections of the source code. It highlights code that has been explicitly recognized as obsolete but is still present in the source code. Obsolescence AD exists in two forms: Current Obsolescence, referring to sections of code that are already obsolete, and Future Obsolescence, referring to sections of code that will become obsolete in the future. Examples of Current and Future obsolescence:

%Obsolescence in software occurs when a technology, component, or system becomes outdated and no longer serves its intended purpose effectively. It implies that newer and more advanced alternatives are available, making the existing solution less relevant or practical. 

%Example:
        \begin{lstlisting}
// Obsoleted method which does nothing.

// This really should be deadcode.

// I think this method can be removed in future versions of JBP.

// This property will be removed in a later release.
    \end{lstlisting}

\textbf{Non-Maintenance:} This refers to the software components or features that are no longer actively maintained or supported due to oversight or uncertainty. It highlights the need for repurposing, refactoring, or removal but does not explicitly confirm that the code is obsolete as in Obsolescence AD. There is a subtle difference between Non-maintenance and Obsolescence AD. Comment which indicates that code that is no longer used is classified as Non-maintenance unless there is an explicit acknowledgement from the developer that it can be removed. Since SAAD is primarily based on comments, tagging unused code as obsolete can be risky as additional validation may be required to assess if the code is referenced elsewhere. Therefore, such code is identified as Non-maintenance AD unless an explicit note for removal is provided. Moreover, unmaintained code can eventually become obsolete or be refactored for reuse, potentially transitioning from the Dormant aging category to the Active aging category. Deprecation is neither unmaintained nor obsolete. It needs to be maintained for a while until the support is required for the feature and eventually it transitions to obsolescence.
%This includes lack of updates, improvements, and bug fixes. When software is not actively maintained, it can become vulnerable to security issues, compatibility problems, and functional limitations over time. 
Examples:
        \begin{lstlisting}
// not used

// the algorithm to follow to perform the check. Currently unused.
    \end{lstlisting}

\subsection{Discussion}
 %Although these systems may not necessarily be flawed in terms of code quality, they often become less efficient and less compatible with new technologies. Dormant SAAD, such as Obsolescence and Non-Maintenance, identifies sections of software that are both inefficient and insecure. 

%Biological aging is categorized into two types, namely, Primary and Secondary Aging \citep{troen2003biology}. Software, akin to biological entities, experiences ongoing evolution \citep{parnas1994software}. 
%Active SAAD can enable strategic management of progressive aging or legacy systems.

%Certain software components age and become vulnerable without regular patches \cite{wang2019detecting}. The Open Web Application Security Project (OWASP) recently highlighted \textbf{A06:2021 Vulnerable and Outdated  Components} \footnote{\url{https://owasp.org/www-project-top-ten/}} as one among the top security risk for web apps, citing unawareness of software versions, outdated software, infrequent vulnerability scans, and patch updates. The SAAD taxonomy can complement other evolutionary metrics in ensuring due maintenance effort is

From the developed taxonomy, we observe that the evolutionary software aging is broadly categorized into two categories, Active and Dormant aging. The proposed taxonomy reflects different types of evolutionary software aging and facilitates the systematic identification of various debts associated with software aging. 

\begin{figure*}
    \centering
    \includegraphics[width=0.95\textwidth]{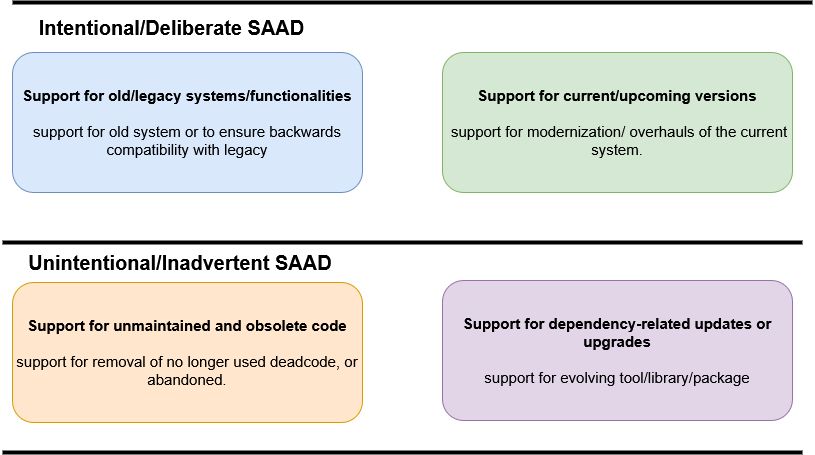}
    \caption{SAAD Quadrants}
    \label{fig:ad_quadrants}
\end{figure*}

Figure \ref{fig:ad_quadrants} indicates the delineation of the different types of AD conceptually inspired from TD quadrants\footnote{\url{https://martinfowler.com/bliki/TechnicalDebtQuadrant.html}}. 

%Although the types of AD conceptually unique, in real time, there can be overlap between AD types. For example, deprecated code can still be kept for backwards compatibility which would warrant to account that code under both legacy and deprecation. 
We differentiate the types of AD based on whether the need to support AD instances arises due to intentional choices or as an unintended consequence of software evolution. Due to the deliberate support required for legacy systems or deprecated functionalities, Legacy and Backwards Compatibility AD and Deprecation AD fall under quadrant 1. Aging Maintenance AD, which represents the ongoing and deliberate effort to modernize or refactor actively used code, falls under quadrant 2. Aging source code transitioning into unused or obsolete is unintentional; therefore, Non-Maintenance and Obsolescence AD fall under quadrant 3. Source code adapting to new updates or upgrades from dependencies is not an intentional adaptation, making Updates or Upgrades fall under quadrant 4.

Legacy code is driven by the need to maintain old systems that are crucial for business operations. This makes legacy code maintenance and compatibility with the latest technologies a result of business considerations rather than purely technical shortcuts. Treating legacy code maintenance overhead as a type of Legacy and Backwards Compatibility AD, which is a type of Active SAAD, accurately reflects the true nature of the maintenance burden. It highlights an aging-related effort to maintain compatibility rather than a sub-optimal solution delivered to meet immediate deadlines. 

Dead code removal \citep{romano2024folklore, caivano2021exploratory, romano2018multi} significantly improves the internal code structure and reduces compile time. We categorize dead code removal under the Dormant AD category as Obsolescence AD. Similarly, stale (i.e., unused) code that developers explicitly acknowledge in comments falls under the Dormant AD category as Non-Maintenance AD. In some cases, unattended deprecated code can create bugs, thereby increasing maintenance overhead \citep{li2018characterising}. We categorize the pending maintenance effort associated with deprecation (aged-out) features or functionalities refactoring or removal as Deprecation AD. To the best of our knowledge, our taxonomy is the first to explicitly identify these pending maintenance efforts exclusively as aging-related debt, whereas conventional TD implicitly view them as delayed/deferred maintenance work \citep{shackleton2023dead, sawant2018reaction}. 

Delineating the maintenance effort associated with obsolete, stale, and deprecated code as AD can help track these debts effectively, thereby giving exclusive focus to the AD and TD items. Consequently, the software maintenance team can better track, plan, and address these issues in a timely manner.

Overall, Active SAAD represents the debt associated with progressive software aging. Dormant SAAD represents the maintenance burden associated with unmaintained or abandoned code that has aged over time and can potentially be detrimental to system maintainability and understanding. The developed taxonomy comprehensively captures the temporal effects of software aging, as reflected in source code comments, offering practitioners a metric to prioritize different types of debts according to their projects. Another interesting possibility is that the proposed SAAD taxonomy, which captures various aspects of software aging reflected in source code comments, can help detect the aging-specific entropy introduced in the software and how it affects the overall software entropy.

%Deprecated code explicitly indicates that a feature or functionality has aged and is outdated. There exists an imminent pending maintenance effort associated with the safe removal of deprecated code and rerouting any dependencies that exist on that code. Sometimes, deprecated code creates bugs, further increasing the maintenance overhead \citep{li2018characterising} and highlighting the need to explicitly identify and track the debt associated with deprecated code. To the best of our knowledge, our taxonomy is the first to identify the debt associated with aging deprecated code maintenance as Deprecation AD.

 %By tracking each type of AD over time, we can if Active aging measure how the entropy of software that reflects the temporal degradation of its structure changes.
%Furthermore, this taxonomy offers a structured approach to deciding whether to update, maintain, or retire these systems. It 
%Another notable finding is that deprecated features and functionalities in the source code are often overlooked in the conventional TD framework, as they do not correspond to the compromises implemented to meet immediate deadlines; rather, they represent explicit software aging-related debt that is no longer relevant. Our taxonomy accurately captures Deprecation AD as a type of Dormant AD. 

%Many systems become integral to operations over extended periods, even as the underlying technology becomes outdated.  The Deprecation SAAD, when used in conjunction with other techniques, can aid in improved code cleanup. 

\begin{tcolorbox}
\textbf{Implications:} %The taxonomy guides practitioners in managing software aging complexities, focusing on dormant components for proactive maintenance. Researchers can explore aging challenges and the need for improved detection methods. 
For practitioners, the proposed taxonomy offers a comprehensive taxonomy to create proactive strategies for detecting and maintaining aging software components. 
%The taxonomy helps categorize various types of aging-related issues, enabling practitioners to more easily identify, understand, and systematically address these issues. 
For researchers, the taxonomy presents a wide range of opportunities for academic exploration of the challenges of software aging.
%The taxonomy not only offers a deeper understanding of the unique challenges of software aging but also highlights the importance of improving aging detection methods and understanding the broader implications of software aging.
\end{tcolorbox}

\section{RQ3: SAAD Prevalence}
\label{aging_prevalence}
\subsection{Motivation}

RQ3: What is the prevalence of SAAD in OSS repositories?

After detecting and categorizing AD through source code comments in OSS repositories, it is essential to understand how prevalent SAAD and its categories are in these repositories. To accomplish this, we evaluate the prevalence of SAAD and its categories in  the PENTACET corpus, which contains over 9,000 OSS repositories with active commit histories.  

In addition to the SAAD patterns identified in Section \ref{b_ad_patterns}, there is a known pattern for Deprecation AD that exists in Java source code comments, which must be included for assessing the prevalence of SAAD in OSS repositories. In Java, the concept of software aging was initially addressed with the introduction of the `@deprecated' Javadoc tag in JDK 1.1, but Java 1.5 introduced the `@Deprecated' annotation to provide compiler-level warnings for developers using deprecated artifacts\footnote{\url{https://openjdk.org/jeps/277}}. In the following text, we refer to the Javadoc tag and annotation as `Existing\_Aging\_Feature', i.e., lower-cased `@deprecated' feature. The `Existing\_Aging\_Feature' was intentionally not included during SAAD exploration in Section \ref{aad_exploration} as adding a known feature would not contribute but only add noise to the code context evaluated SAAD pattern extraction step. Although `@Deprecated' annotations can appear in the source code without an accompanying comment, we exclude such cases from this study since our focus is limited to studying aging-related debt from source code comments. Therefore, this study covers `Existing\_Aging\_Feature' included in the source code comments only. There are 46,147 comments with `Existing\_Aging\_Feature' in PENTACET corpus. %This leads to an intriguing question: How much aging-related debt might we be missing beyond the proposed Deprecation AD, conventionally identified as aged code using `Existing\_Aging\_Feature', overlooking the pending maintenance effort? 

Overall, we quantify SAAD, including both `Existing\_Aging\_Feature' and the determined SAAD patterns, to have an accurate understanding of its prevalence. In addition, we evaluate the dominant AD category in OSS repositories as observed in the taxonomy shown in Figure \ref{fig:taxonomy}. %Importance of software aging has been recognized by software engineering community. Clear indication of this is that for Java-language since version 5
%SonarQube, one of the widely used static code analysis tool, aids software maintenance by identifying bugs, vulnerabilities, and code smells to refine code quality. Notably, while SonarQube leverages the 
%offers compiler level support `@deprecated' feature that works an aging indicator.
%Our SAAD features don't rely on `Existing\_Aging\_Feature'. 

\subsection{Approach}
For assessing the prevalence of SAAD, we utilize both the gold standard and silver standard SAAD datasets from RQ 1. The objective of using a machine-annotated silver standard dataset is to triangulate the findings of SAAD prevalence from the gold standard dataset. 

To evaluate the prevalence of SAAD (RQ 3), we first quantitatively assess the proportion of source code comments classified as SAAD using the PENTACET corpus. For the gold standard dataset, comprising 2,562 SAAD comments from Section \ref{aad_exploration}, the first author manually reads the comment to classify each SAAD comment into its respective SAAD type identified in the SAAD Taxonomy \ref{fig:taxonomy}. For the silver standard dataset, comprising 35,630 SAAD comments, the first author utilizes the code context evaluated SAAD patterns (used to identify the AD types evaluated by the second author in Section \ref{aging_taxonomy}) to perform pattern-based type classification. For the gold standard SAAD dataset, if an SAAD comment reflects multiple SAAD types, it is classified under multiple AD types (i.e., it is counted as a unique AD for each AD type identified in the comment, a similar approach as in \citep{obrien202223}). For the silver standard dataset, if an SAAD pattern corresponding to a particular SAAD type (for example, Legacy \& Backwards Compatibility AD) is present in an SAAD comment, that comment is classified as Legacy \& Backwards Compatibility AD. If there are overlapping SAAD patterns, that is, a comment containing SAAD patterns corresponding to more than one AD type, then that SAAD comment is counted in each of those categories. For example, consider the following comment that contains the patterns `this is useless' and `provide backwards compatibility' is counted under both `Non-Maintenance' and `Legacy \& Backwards Compatibility' AD types.

\begin{lstlisting}
    /* this is useless except to provide backwards compatibility in 
    * phi_convict_threshold because everyone seems pretty 
    * accustomed to the default of 8 */
\end{lstlisting}

We group the identified types of AD into `Active' or `Dormant' AD categories. To ascertain if there is any statistically significant difference between Active and Dormant AD categories in OSS repositories, we evaluate the following hypothesis,

\textbf{Null Hypothesis ($H_0$)}: No significant difference exists between the prevalence of Dormant SAAD comments and Active SAAD comments in OSS repositories.

\textbf{Alternative Hypothesis ($H_\alpha$)}: A significant difference exists between the prevalence of Dormant SAAD comments and Active SAAD comments in OSS repositories.

%Using the outcomes from our previous qualitative analysis of source code comments and the Aging Debt Taxonomy depicted in \ref{fig:taxonomy}, we categorize the identified different types of SAAD comments into `Active' or `Dormant' aging debt. Following this classification, our goal is to ascertain if Dormant Aging is more prevalent in OSS software repositories compared to Active Aging.

%This evaluation provides insight into software robustness, highlighting whether recognized software aging issues remain latent ('Dormant') or currently impact software performance ('Active'). 
Given its capability to handle data without assuming a normal distribution, we employ the Wilcoxon signed-rank test, a non-parametric statistical method, to assess our hypothesis. We calculate the effect size, denoted as \( r \), using the formula:

\[
r = \frac{Z}{\sqrt{N}}
\]

where \( Z \) is the standardized test statistic derived from the Wilcoxon signed-rank test, and \( N \) represents the total number of non-zero differences between the paired observations of Dormant SAAD and Active SAAD comments in OSS repositories. 

This effect size \( r \) is interpreted as a rank-biserial correlation \citep{tomczak2014need}, providing a measure of the strength of the relationship between the two matched groups. The value of \( r \) ranges from -1 to 1, where a positive \( r \) indicates that the median of the first group is typically higher than that of the second group, and a negative \( r \) suggests the opposite. Regarding the magnitude of the effect size, similar to other correlation coefficients, the closer the absolute value \( |r| \) is to 1, the stronger the effect. Commonly, magnitudes of \( |r| \) are interpreted as in \citep{cohen2013statistical}: 0.1 $<$ \( |r| \) $<$ 0.3 is considered small, 0.3 $<$ \( |r| \) $<$ 0.5 as medium, and \( |r| \) $>$ 0.5 as large.

%Furthermore, to gain a deeper understanding of SAAD's value, we identify all natural language comment instances of the `Existing\_Aging\_Feature' within the same PENTACET corpus. This will help us determine how many new instances of AD can be identified by SAAD, beyond those covered by the existing `@deprecated' Javadoc tag or the `@Deprecated' annotation.

\subsection{Results}

%Table \ref{tab:active_dormant_statistic} and 
Tables \ref{tab:ad_prevalence} present the prevalence of SAAD in relation to the number of projects and the total number of natural language comments from the PENTACET corpus. Conversely, Table \ref{tab:active_vs_dormant_ad} breaks down the various types and categories of AD among the SAAD mentioned in Table \ref{tab:ad_prevalence}. Together, these tables offer complementary insights into the prevalence and nature of AD in OSS repositories. Table \ref{tab:active_dormant_hypothesis} shows the results of hypothesis evaluation for assessing the dominant AD category using the Wilcoxon Signed-Rank. %The average percentage of Active SAAD across the projects is 40.93\%. The median value of Active SAAD, at 33.33\%, indicates that while most projects have a lower prevalence of Active SAAD, a few projects with very high Active SAAD values shift the mean upwards. Similarly, the average percentage of Dormant SAAD is 59.07\%, indicating that more than half of the SAAD across projects falls into the Dormant category. The median value of Dormant SAAD, at 66.67\%, suggests that most projects contain Dormant SAAD instances closer to this higher median value.

\begin{comment}
\begin{table}[htbp]
\centering
  \caption{Active VS Dormant SAAD Prevalence: Descriptive Statistics}
  \label{tab:active_dormant_statistic}
  \begin{tabular}{ccccc}
    \toprule
    \multirow{2}{*}{\makecell{\textbf{PENTACET} \\ \# \textbf{SAAD afflicted} \\ \textbf{Projects}}} & \multicolumn{2}{c}{\makecell{\textbf{Active} \\ \textbf{SAAD}}} & \multicolumn{2}{c}{\makecell{\textbf{Dormant} \\ \textbf{SAAD}}}  \\
    \cmidrule(lr){2-3}
    \cmidrule(lr){4-5}
    & \makecell{\textbf{Mean} \\ } & \makecell{\textbf{Median} \\ } & \makecell{\textbf{Mean}} & \makecell{\textbf{Median}} \\
    \midrule
    1774 & 40.93 & 33.33 & 59.07 & 66.67 \\
    \bottomrule
  \end{tabular}
\end{table}
\end{comment}

\begin{table}[htbp]
\caption{SAAD Prevalence}
\label{tab:ad_prevalence}
\resizebox{\columnwidth}{!}{%
\begin{tabular}{ccccc}
\hline
\textbf{\begin{tabular}[c]{@{}c@{}}SAAD\\ Dataset\end{tabular}} &
  \textbf{\begin{tabular}[c]{@{}c@{}}Project \&\\ Comment Counts\end{tabular}} &
  \textbf{\begin{tabular}[c]{@{}c@{}}Total in\\ PENTACET\end{tabular}} &
  \textbf{\begin{tabular}[c]{@{}c@{}}\# with\\ SAAD\end{tabular}} &
  \textbf{\begin{tabular}[c]{@{}c@{}}\% Afflicted\\ with SAAD\end{tabular}} \\ \hline
\multirow{2}{*}{\begin{tabular}[c]{@{}c@{}}Gold Standard\\ Dataset\end{tabular}}   & \# of Projects    & 9094       & 1,957  & 21.52 \\ \cline{2-5} 
                                                                                   & \# of NL Comments & 16,153,942 & 48,709 & 0.3   \\ \hline
\multirow{2}{*}{\begin{tabular}[c]{@{}c@{}}Silver Standard\\ Dataset\end{tabular}} & \# of Projects    & 9094       & 3,053  & 33.57 \\ \cline{2-5} 
                                                                                   & \# of NL Comments & 16,153,942 & 81,777 & 0.5   \\ \hline
\end{tabular}%
}
\end{table}

\begin{table}[htbp]
\caption{SAAD Prevalence Across Types \& Categories}
\label{tab:active_vs_dormant_ad}
\resizebox{\columnwidth}{!}{%
\begin{tabular}{cccccc}
\hline
\textbf{\begin{tabular}[c]{@{}c@{}}SAAD Dataset\\ Type\end{tabular}} &
  \textbf{\begin{tabular}[c]{@{}c@{}}SAAD\\ Category\end{tabular}} &
  \textbf{\begin{tabular}[c]{@{}c@{}}SAAD\\ Type\end{tabular}} &
  \textbf{\begin{tabular}[c]{@{}c@{}}\# of\\ Instances\end{tabular}} &
  \textbf{\begin{tabular}[c]{@{}c@{}}SAAD Type\\ \% Affliction\end{tabular}} &
  \textbf{\begin{tabular}[c]{@{}c@{}}SAAD Category\\ \% Affliction\end{tabular}} \\ \hline
\multirow{9}{*}{\textbf{\begin{tabular}[c]{@{}c@{}}Gold Standard \\ SAAD\\ Dataset\\ (Fully Manually\\ Verified)\end{tabular}}} &
  \multirow{3}{*}{\textbf{\begin{tabular}[c]{@{}c@{}}Active\\ SAAD\end{tabular}}} &
  \textbf{\begin{tabular}[c]{@{}c@{}}Aging\\ Maintenance\end{tabular}} &
  122 &
  0.25 &
  \multirow{3}{*}{3.47} \\ \cline{3-5}
 &
   &
  \textbf{\begin{tabular}[c]{@{}c@{}}Legacy Backwards\\ Compatibility\end{tabular}} &
  1,459 &
  2.97 &
   \\ \cline{3-5}
 &
   &
  \textbf{Updates-Upgrades} &
  121 &
  0.25 &
   \\ \cline{2-6} 
 &
  \multirow{6}{*}{\textbf{\begin{tabular}[c]{@{}c@{}}Dormant\\ SAAD\end{tabular}}} &
  \textbf{\begin{tabular}[c]{@{}c@{}}Current Deprecation\\ (from SAAD patterns)\end{tabular}} &
  540 &
  1.10 &
  \multirow{6}{*}{\textbf{96.53}} \\ \cline{3-5}
 &
   &
  \textbf{\begin{tabular}[c]{@{}c@{}}Current Deprecation\\ (from Existing\_Aging\_Feature)\end{tabular}} &
  46,147 &
  94.01 &
   \\ \cline{3-5}
 &
   &
  \textbf{Future Deprecation} &
  25 &
  0.05 &
   \\ \cline{3-5}
 &
   &
  \textbf{Non-Maintenance} &
  182 &
  0.37 &
   \\ \cline{3-5}
 &
   &
  \textbf{Current Obsolescence} &
  128 &
  0.26 &
   \\ \cline{3-5}
 &
   &
  \textbf{Future Obsolescence} &
  362 &
  0.74 &
   \\ \hline
\multirow{9}{*}{\textbf{\begin{tabular}[c]{@{}c@{}}Silver Standard\\ SAAD\\ Dataset\\ (SAAD pattern-\\ based)\end{tabular}}} &
  \multirow{3}{*}{\textbf{\begin{tabular}[c]{@{}c@{}}Active\\ SAAD\end{tabular}}} &
  \textbf{\begin{tabular}[c]{@{}c@{}}Aging\\ Maintenance\end{tabular}} &
  1,873 &
  2.27 &
  \multirow{3}{*}{18.74} \\ \cline{3-5}
 &
   &
  \textbf{\begin{tabular}[c]{@{}c@{}}Legacy-\\ Backwards Compatibility\end{tabular}} &
  13,296 &
  16.09 &
   \\ \cline{3-5}
 &
   &
  \textbf{Updates-Upgrades} &
  315 &
  0.38 &
   \\ \cline{2-6} 
 &
  \multirow{6}{*}{\textbf{\begin{tabular}[c]{@{}c@{}}Dormant\\ SAAD\end{tabular}}} &
  \textbf{\begin{tabular}[c]{@{}c@{}}Current Deprecation\\ (from SAAD patterns)\end{tabular}} &
  14,379 &
  17.40 &
  \multirow{6}{*}{\textbf{81.26}} \\ \cline{3-5}
 &
   &
  \textbf{\begin{tabular}[c]{@{}c@{}}Current Deprecation\\ (from Existing\_Aging\_Feature)\end{tabular}} &
  46,147 &
  55.85 &
   \\ \cline{3-5}
 &
   &
  \textbf{Future Deprecation} &
  67 &
  0.08 &
   \\ \cline{3-5}
 &
   &
  \textbf{Non-Maintenance} &
  4,843 &
  5.86 &
   \\ \cline{3-5}
 &
   &
  \textbf{Current Obsolescence} &
  922 &
  1.12 &
   \\ \cline{3-5}
 &
   &
  \textbf{Future Obsolescence} &
  782 &
  0.95 &
   \\ \hline
\end{tabular}%
}

\end{table}

Table \ref{tab:ad_prevalence} shows that among the 9,094 OSS projects in the PENTACET corpus, SAAD is present in 21.52\% of the projects for the gold standard (1,957 projects) and 33.57\% for the silver standard dataset (3,053 projects), respectively. In terms of the total number of natural language comments in the PENTACET corpus, SAAD appears in 0.3\% of the comments for the gold standard (48,709 instances) and 0.5\% for the silver standard dataset (81,777 instances), respectively.

For the gold standard dataset, Table \ref{tab:active_vs_dormant_ad} shows that Aging Maintenance and Updates \& Upgrades are the least represented within the Active SAAD category, with 122 and 121 instances, respectively. They are followed by Legacy \& Backwards Compatibility with 1,459 SAAD instances. Within the Dormant AD category, Deprecation AD stands out prominently. Specifically, Current Deprecation results in 46,712 SAAD instances with 46,147 from `Existing\_Aging\_Feature' and 540 from SAAD patterns. The Future Deprecation AD accounts for 25 SAAD instances. Obsolescence AD trails behind Deprecation AD, with 490 SAAD instances (128 for Current Obsolescence and 362 for Future Obsolescence AD, respectively). The Non-Maintenance AD accounts for the least, with 182 SAAD instances. In summary, for the gold standard dataset, Active SAAD types contribute 3.47\% and Dormant SAAD accounts for 96.53\% of the total SAAD instances.

The silver standard dataset displays a similar trend to the gold standard dataset, with the Dormant AD category accounting for 81.26\% of SAAD instances. It surpasses the Active AD category, which accounts for 18.74\%. Within the Active AD category, Legacy \& Backwards Compatibility AD is the most prominent, with 13,296 SAAD instances. This is followed by Aging Maintenance AD (1,873 instances) and Updates-Upgrades AD (315 instances). In the Dormant AD category, Deprecation AD is the most prevalent, with 60,973 SAAD instances. Deprecation AD is followed by Non-Maintenance AD (4,843 SAAD instances) and Obsolescence AD (1,704 SAAD instances). The Dormant AD in the silver standard is smaller than the gold standard by 15.27\%. This is primarily due to the increase in Legacy and Backwards Compatibility AD and Aging Maintenance AD of the Active AD category.
%The silver standard dataset shows a similar trend to the gold standard dataset, with the Dormant AD category accounting for 81.26\% of total SAAD instances. It dominates the Active AD category, which accounts for 18.74\%. In the Active AD category, Legacy \& Backwards Compatibility AD is the most prominent, with 13,296 SAAD instances. It is followed by Aging Maintenance AD (1,873 instances) and Updates-Upgrades AD (315 instances). Similarly, in the Dormant AD category, Deprecation AD is the most prominent, accounting for 60,973 SAAD instances. Non-Maintenance AD follows with 4,843 SAAD instances, and Obsolescence AD contributes 1,704 SAAD instances. The overall trend of the silver standard aligns with the gold standard data, indicating Dormant AD being more prevalent than Active AD. However, the Dormant AD in the silver standard is relatively smaller than gold standard by 15.27\%. It is primarily due to the increase in Legacy and Backwards Compatibility AD and Aging Maintenance AD of the Active Aging instances.

From Table \ref{tab:active_dormant_hypothesis}, the extremely small p-value (less than 0.001) obtained from the Wilcoxon-signed rank test based on the SAAD prevalence from 1,957 projects (gold standard) and 3,053 projects (silver standard) allows us to confidently reject the Null Hypothesis $H_0$, which states that there is no significant difference between the prevalence of Dormant SAAD comments and Active SAAD comments in OSS repositories. The extremely low p-value supports the Alternative Hypothesis $H_\alpha$, which affirms that a significant difference exists between the prevalence of Dormant AD comments and Active AD comments in both gold and silver standard SAAD data. In addition, an effect size of \(r = -0.692\) implies a large negative effect, indicating that the prevalence of Active SAAD is significantly lower than that of Dormant SAAD in OSS repositories. This suggests that the practical difference in SAAD prevalence is substantial. A similar trend is observed in the silver standard SAAD data, which shows a medium negative effect, again indicating that Dormant SAAD is more common than Active SAAD. However, despite the statistically significant trend being similar across both datasets, the difference in magnitude between the gold and silver standard SAAD data is primarily due to the larger number of projects in the silver standard, with relatively more Active AD instances than the gold standard. 

Another finding is that the `Existing\_Aging\_Feature' of Deprecation AD is the main contributor of Dormant SAAD in both gold and silver standard datasets. The `Existing\_Aging\_Feature' contributes 94.01\% and 55.85\% to SAAD comments across gold and silver standard datasets, respectively. This shows that Deprecation AD is the most dominant AD across all AD types in OSS repositories.

\begin{table}[]
\caption{Active VS Dormant SAAD Prevalence: Hypothesis Evaluation (Wilcoxon Test)}
\label{tab:active_dormant_hypothesis}
\resizebox{\columnwidth}{!}{%
\begin{tabular}{cccccc}
\hline
\textbf{\begin{tabular}[c]{@{}c@{}}SAAD\\ Dataset\end{tabular}} &
  \textbf{\begin{tabular}[c]{@{}c@{}}PENTACET \\ Projects with SAAD \\ Comments\end{tabular}} &
  \textbf{\begin{tabular}[c]{@{}c@{}}Wilcoxon\\ Statistic\\ (W)\end{tabular}} &
  \textbf{p-value} &
  \textbf{\begin{tabular}[c]{@{}c@{}}Effect Size\\ Estimate (r)\end{tabular}} &
  \textbf{\begin{tabular}[c]{@{}c@{}}Magnitude\\ $|$r$|$\end{tabular}} \\ \hline
\textbf{\begin{tabular}[c]{@{}c@{}}Gold\\ Standard\\ SAAD\end{tabular}} &
  1957 &
  181312 &
  $<$0.001 &
  -0.692 &
  Large \\ \hline
\textbf{\begin{tabular}[c]{@{}c@{}}Silver\\ Standard\\ SAAD\end{tabular}} &
  3053 &
  1161052 &
  $<$0.001 &
  -0.389 &
  Medium \\ \hline
\end{tabular}%
}
\end{table}

\subsection{Discussion}
There is clear, statistically significant evidence showing that Dormant AD is more prevalent in OSS repositories than Active AD in gold standard data. The higher prevalence of Dormant AD in both datasets indicates that a significant amount of SAAD in OSS repositories is related to the dominant Deprecation AD of the Dormant AD category. This prevalence makes it evident that developers often leave deprecation notes for the aged-out features or functionalities that are no longer in active use, with a pending maintenance effort to temporarily maintain or safely remove deprecated code and its references while ensuring the current business flow of the software is not affected.

The Dormant SAAD with 96.53\% of SAAD instances in the gold standard data indicates a higher presence of outdated software components in OSS repositories. This reflects the rapidly evolving nature of software OSS repositories. On the other hand, the prevalence of Legacy \& Backwards-Compatibility SAAD that makes up 2.97\% of Active SAAD in gold standard SAAD, shows there are challenges in maintaining legacy and compatibility with older software versions in OSS repositories.

The triangulation of the AD prevalence between the manually verified (gold standard) and pattern-based (silver standard) data reinforces the credibility of the findings. Despite minor differences in magnitude, likely due to the larger number of projects in the silver standard data with more Active AD instances, the statistically significant trends underscore the reliability of the prevalence patterns. Overall, the data triangulation confirms that Dormant AD is more prevalent than Active AD in OSS repositories. From a practical utility perspective, these findings indicate that OSS repositories have more dormant AD in the form of outdated code for removal or deprecation refactoring, which is often overlooked.

\begin{tcolorbox}
\textbf{Implications:} Practitioners often encounter Deprecation SAAD in OSS repositories, followed by compatibility challenges for legacy systems which calls for different strategies of software maintenance. Researchers have the opportunity to explore and delve into static and dynamic analysis of software metrics, enhancing their understanding of Active and Dormant AD in detail and its impact on software maintenance and overall software health. %use the AD taxonomy to uncover aging instances not captured by traditional '@deprecated' methods, indicating potential improvements in AD detection techniques. %For researchers, the introduced AD taxonomy identifies aging instances often overlooked by the traditional `@deprecated' feature, suggesting room for enhancing AD detection techniques.
\end{tcolorbox}

\section{Related Work}
\label{rw}
\subsection{Runtime Aging}
Past studies have focused on detecting the temporal degradation of software during runtime through multiple metrics and dimensions. A regression model, utilizing response time and failure rate, is proposed as a comprehensive metric to evaluate software aging \citep{lei2010comprehensive}. Key metrics include memory utilization \citep{shereshevsky2003software}, resource utilization and depletion \citep{garg1998methodology, castelli2001proactive, grottke2006analysis, zhao2013ensuring}, performance metrics such as response time \citep{avritzer2006performance}, and data traffic between software systems \citep{avritzer1997monitoring}. Newer metrics such as Fluency, which factors in the active state of mobile software, and User Experience Degree, which accounts for user experience across different software states, are also proposed for measuring software aging. Additionally, a methodology for the runtime evaluation of different software versions to detect memory leaks, as well as a metric for assessing the severity of software aging, are introduced \citep{zheng2012advanced}.

\subsection{Evolutionary Aging}
Previous research has examined various aspects of software evolution over time. Within the context of C++, a data model was developed to detect dead code \citep{chen1998c++}. Historically, the detection and elimination of dead code have been compiler-driven optimization processes \citep{torczon2007engineering}. Piranha, a recent tool, tackles feature debt and dead code by identifying and removing outdated and unused flags \citep{ramanathan2020piranha}. In JavaScript, the JSNose tool employs dynamic analysis metrics, including execution count and code reachability, to detect and recommend unused code for refactoring \citep{fard2013jsnose}. An automatic program transformation tool was introduced to address collateral evolutions in system software, particularly for Linux device drivers, ensuring that developers updating libraries also make the necessary changes in drivers maintained outside the kernel source tree \citep{padioleau2008documenting}. The Record and Replay approach, utilized in integrated development environments (IDEs), records and replays refactoring events within code libraries to assist in detecting software evolution \citep{henkel2005catchup}. GitHub version history data has been leveraged to assess the co-evolution behavior of the Android API and dependent applications \citep{mcdonnell2013empirical}. The identification of outdated and faulty third-party libraries has been achieved using a smart alerting system linked to an external database \citep{wang2020empirical}. Method inlining, a technique involving replacing calls to a deprecated method (in Java either with @Deprecated annotation or with @deprecated tag) with the actual code contained within the method, has also been documented \citep{perkins2005automatically}. Unlike the proposed AD taxonomy that focuses on aging-related debt, a taxonomy of deprecation reasons is established \citep{sawant2018features}, highlighting the reasons for deprecating API features conveyed in the deprecation message. Most recently, SCARF has been introduced as a fully automated framework for detecting and deprecating unused source code and datasets using runtime metrics \citep{shackleton2023dead}. To the best of our knowledge, our study is the first to develop a taxonomy by studying aging-related debt from source code comments.

In contrast, our study focuses on evolutionary software aging by introducing a potentially measurable metric through source code comments known as SAAD. This metric incorporates a comprehensive taxonomy addressing various issues related to evolutionary software aging, including dead code, code refactoring, identification of outdated components, deprecation reasons, method inlining, and more. Our novel approach, SAAD, adopts a developer-centric perspective, classifying these issues into Active and Dormant aging categories. SAAD analyzes software artifacts through natural language source code comments, making it programming language-agnostic. This method complements existing evolutionary and runtime metrics on aging.

\section{Threats to Validity}
\label{validity}
\subsection{Internal Validity}
One primary concern in RQ 1 (described in Section \ref{aad_exploration}) involves using the Sense2Vec AI model to extrapolate software aging-related features, which relies on training data derived from software engineering comments from the Reddit platform. It is essential to note that prior studies \citep{sheikhaei2024empirical, sridharan2023pentacet} have employed Sense2Vec to detect debt in software. In addition, we enhanced the robustness of the software aging-related features by implementing a multi-level manual review process to determine related features. We identified 2 core features from seminal software aging-related research \citep{parnas1994software} and 6 ad hoc features that represent the software aging theme and complement the core features as seed features. We input the seed features related to software aging and employed a filtering criterion to assess whether a text feature implies software aging. This criterion is aimed at including all potential features related to software aging without any preconceived bias. After the manual analysis adhering to the filtering criterion, the resulting set of aging-related text features undergoes another round of validation by the second author to filter out features that are not related to software aging. Through these steps, we ensured a systematic approach and removed bias in identifying software aging-related features. 

Another potential concern is the bias associated with the manual labeling of source code comments. To mitigate this risk, we included the context of each comment during the manual analysis and conducted an inter-rater agreement exercise with the third author, who was not involved in any previous steps. Although the IRA score of 0.695 is not excellent in terms of Cohen's Kappa interpretation \citep{emam1999benchmarking}, it falls within the good range, reaffirming that the manual labeling is unbiased, and indicating that there exists substantial evidence that the annotated comments are truly SAAD comment.  
\subsection{Construct Validity} 
The comprehensiveness of the developed SAAD taxonomy presents a potential threat to construct validity in that it may not cover all facets of software aging. To mitigate this risk, first, we utilize the 145 systematically identified software aging–related features that cover the different facets of software aging to annotate 16.15 million source code comments. Next, we perform a subsequent manual analysis with code context to determine SAAD comments and extract SAAD text patterns, ensuring that at least 95\% of unique SAAD patterns are captured, resulting in a robust set of code context–evaluated SAAD patterns. These steps ensure the comprehensiveness of the developed SAAD taxonomy based on text features from source code comments. Additionally, this is the first attempt to develop a taxonomy based on software aging-related features from source code comments.  Future works can expand our taxonomy by incorporating other dynamic software aging-related metrics.
\subsection{External Validity}
While our large corpus strengthens our ability to generalize findings across OSS projects, the exclusive utilization of the PENTACET corpus in this study means our findings primarily apply to Java OSS repositories. This poses a limitation when considering other programming languages. Moreover, software practices, comment styles, and aging-related issues may vary significantly across different languages and may entirely differ in proprietary software environments. Future research might expand on our methodology by utilizing our gold standard dataset to train AI models in diverse corpora, thereby enhancing the generalizability of SAAD across different programming languages and environments (OSS/proprietary).

% or by incorporating proprietary software datasets to enhance the scope of generalizability.

\section{Conclusion}
\label{conclusion}
Our study makes four main contributions. Firstly, we provide a novel view of software aging by analyzing a wide spectrum of prior works and by introducing high-level aging categories of Evolutionary Aging, which refers to aging due to software evolution, and Runtime Aging, which refers to an increasing number of reliability issues as software runtime increases; see Section \ref{sec:two-views}. We extend the evolutionary view of software aging by introducing AD and differentiating it from TD; see Section \ref{sec:AD_vs_TD}. AD refers to the cumulative challenges and issues that software systems face as they evolve and mature over time.
We investigate AD expressed in the source code comments through qualitative and quantitative analysis of source code comments.

Secondly, we develop a method (Figure \ref{exploring_aad}) to detect AD empirically occurring in source code comments, terming such instances as SAAD comments. Using this method, we create 399 code context evaluated SAAD text patterns and a gold standard dataset for SAAD comments comprising 2562 SAAD comments.

Thirdly, leveraging the 399 SAAD text patterns, we conduct axial and selective coding to construct a comprehensive SAAD taxonomy (Figure \ref{fig:taxonomy}). This taxonomy classifies SAAD into `Active' and `Dormant' categories, each with various sub-types. 

Our fourth contribution involves the prevalence analysis of SAAD using the developed taxonomy in OSS repositories. Although this taxonomy may not have exhaustively characterized every facet of software aging, it represents the first attempt to capture various dimensions of software aging through source code comments. Our investigation, which used 399 SAAD patterns, identified over 45,000 SAAD instances, indicating its practical significance. Notably, our findings underscore the prevalence of Dormant AD, which constitutes 96.53\% of SAAD instances in our gold standard SAAD dataset. Furthermore, data triangulation with the silver standard SAAD dataset shows that the Dormant AD category accounted for 81.36\%. This corroborates the finding that Dormant SAAD is more prevalent in OSS repositories.
%and affirms the accuracy and generalizability of SAAD patterns used for annotating silver standard dataset. 
%and SAAD data.

%Our fourth contribution involves an analysis of Open Source Software (OSS) using this SAAD taxonomy. Although this taxonomy can not be exhaustive in characterizing the different facets of software aging, this is the first attempt in capturing different dimensions of software aging through source code comments. Our investigation, with a limited set of SAAD patterns equates to over 45,000 SAAD instances, signifying practical significance. Notably, our findings underscore the prevalence of Dormant Aging Debt, constituting 96.53\% of SAAD instances in our gold standard SAAD dataset. Evaluating the extracted code context evaluated SAAD patterns through data triangulation using the silver standard SAAD dataset with 81.36\% of dormant AD category further reaffirms the accuracy of SAAD patterns and SAAD data.
%, reveals SAAD in $\approx$0.1\% of comments from a vast pool of 16 million natural language comments. While this percentage might seem modest, it 

%While AD and TD have distinct origins, unique characteristics with some overlaps treating AD as a type of TD for measurement and management purposes can provide practical benefits. This approach allows for a more integrated, efficient, and systematic handling of the various issues that affect software over its lifecycle. However, it is crucial to recognize the specific nuances of AD that warrant targeted strategies and relevant prioritization.

We identify numerous opportunities for future research in the area of evolutionary software aging. Our taxonomy, which is based on SAAD comments, could be extended to study other aspects of evolutionary software aging, such as technical lag and dead code detection. An important area for future work would be to explore the connections between evolutionary aging and software defects. Additionally, employing AI techniques to enhance the accuracy and scope of SAAD and AD detection across multiple programming languages could provide significant benefits, including the ability to identify different types of evolutionary aging such as dead code and TD. Exploring the relationship between SAAD and software security is also crucial, as AD may contribute to security vulnerabilities. Lastly, conducting a longitudinal study of SAAD instances would help verify whether the transitions between dormant and active aging comments are observable through version control changes in software.

\section*{Declarations}
\subsection*{Funding}
Authors acknowledge the financial support of the Academy of Finland through a financial grant (Grant Number 328058). The open access funding of this research work is provided by Academy of Finland and University of Oulu.
\vspace{-0.5cm}
\subsection*{Data Availability}
The SAAD source code comments data along with manually reviewed SAAD features used in our research work is available under a Creative Commons license at \citep{muralisaad2025} to support any future research and ensure the replicability of our findings.
\vspace{-0.5cm}
\subsection*{Conflict of Interests}
The authors declare that there are no conflicts of interest associated with this work, meaning there are no relevant financial or non-financial interests with any entity. Additionally, as recommended by the submission guidelines, we disclose that the second author, Mika Mäntylä, serves on the editorial board of the journal.
\vspace{-0.5cm}
\subsection*{Author Contributions}  Murali Sridharan is responsible for the initial conceptualization, the initial draft, and the final draft after incorporating revisions from external reviewers as well as from the second and third authors. Mika Mäntylä contributed to positioning the concept of AD, conducted a critical review of the entire work, contributed to the final draft, and served as a referee to resolve conflicts during the IRA exercise. Leevi Rantala provided critical feedback throughout this research work and participated in the IRA annotation for the SAAD dataset.
\vspace{-0.5cm}
\subsection*{Ethical Approval}
Not applicable. This work involves exploratory analysis of source code comments from OSS repositories and does not require ethical approval.
\vspace{-0.5cm}
\subsection*{Informed Consent} Not applicable.
\vspace{-0.5cm}
\subsection*{Clinical Trial} Not applicable.

\bibliography{template}
%\printbibliography

\newpage

\appendix
\section{Software Aging related Sense2Vec Extrapolated Features}
\label{sec:appendix_a}

\begin{multicols}{2}
%\begin{enumerate}
\textbf{Direct Aging Features} 
\begin{enumerate}[leftmargin=*]
\item `outdated',
\item `obsolete',
\item `antiquated',
\item `deprecated',
\item `older version',
\item `earlier version',
\item `prior version',
\item `outdated version',
\item `old version',
\item `past version',
\item `previous version',
\item `older model',
\item `previous model',
\item `old model',
\item `old system',
\item `legacy software',
\item `legacy apps',
\item `older software',
\item `old software',
\item `old programs',
\item `legacy~applications',
\item `outdated software',
\item `legacy support',
\item `legacy systems',
\item `legacy code',
\item `old code',
\item `legacy system',
\item `outdated system',
\item `older system',
\item `archaic',
\item `old',
\item `eol',
\item `end-of-life'
\item `unmaintained',
\item `abandoned',
\item `abandoning',
\item `neglected',
\item `deserted',
\item `shelved',
\item `decommissioned',
\item `decommission',
\item `obsoleted'
\item `legacy'
\item `past release'
\item `deadcode'
\item `dead-code'
\end{enumerate}

% Indirect Aging Features
\textbf{Indirect Aging Features}
\begin{enumerate}

\item `redesign',
\item `revamp',
\item `upgrade',
\item `upgradable',
\item `upgradability',
\item `upgrading',
\item `modernize',
\item `modernization',
\item `modernise',
\item `modernisation',
\item `updated',
\item `update',
\item `updating',
\item `newer',
\item `newer version',
\item `newest version',
\item `newest update',
\item `newest patch',
\item `patch',
\item `latest patch',
\item `latest update',
\item `latest build',
\item `new version',
\item `later version',
\item `future versions',
\item `future version',
\item `next version',
\item `next release',
\item `next major release',
\item `new feature',
\item `future release',
\item `next revision',
\item `upcoming version',
\item `new functionality',
\item `latest version',
\item `updated version',
\item `improved version',
\item `upgraded version',
\item `compatible version',
\item `newest model',
\item `newer model',
\item `latest model',
\item `new system',
\item `new program',
\item `new software',
\item `newer software',
\item `security patches',
\item `security updates',
\item `OS upgrades',
\item `backward compatibility',
\item `backwards compatibility',
\item `new implementation',
\item `OS updates',
\item `firmware update',
\item `OTA updates',
\item `software updates',
\item `iOS updates',
\item `Android updates',
\item `software upgrades',
\item `new platform',
\item `newer system',
\item `scrapped',
\item `phased',
\item `phasing',
\item `ditching',
\item `retooled',
\item `retool',
\item `rebuild',
\item `retooling',
\item `overhauling',
\item `overhaul',
\item `nixing',
\item `eliminating',
\item `eliminate',
\item `rid',
\item `ditch',
\item `eradicating',
\item `eradicate',
\item `eradication',
\item `removing',
\item `remove',
\item `delete',
\item `deleting',
\item `purge',
\item `purging',
\item `retrofitted',
\item `retrofit',
\item `axed',
\item `axing',
\item `scrap',
\item `nixed',
\item `ditched',
\item `trashed',
\item `trash',
\item `unnecessary',
\item `unneeded',
\item `unused',
\item `unusable',
\item `useless',
\end{enumerate}
%\end{enumerate}
\end{multicols}

\end{document}